\documentclass{amsart}

\usepackage{amsmath,amsfonts,amssymb,amsthm,graphicx,cite}
\usepackage{youngtab}
\usepackage{mathrsfs}

\newtheorem{lemma}{Lemma}[section]
\newtheorem{prop}{Proposition}[section]
\newtheorem{cor}{Corollary}[section]

\theoremstyle{remark}
\newtheorem{remark}{Remark}[section]

\begin{document}

\title[A unified construction of generalised classical polynomials]{A
unified construction of generalised classical polynomials associated
with operators of Calogero-Sutherland type}

\author{Martin Halln\"as}
\address{SISSA, Via Beirut 2-4, 34151 Trieste TS, Italy}
\email{hallnas@sissa.it}

\author{Edwin Langmann}
\address{Theoretical physics, KTH, S-106 91 Stockholm, Sweden}
\email{langmann@kth.se}

\date{\today}

\begin{abstract}
  In this paper we consider a large class of many-variable polynomials
  which contains generalisations of the classical Hermite, Laguerre,
  Jacobi and Bessel polynomials as special cases, and which occur as
  the polynomial part in the eigenfunctions of Calogero-Sutherland
  type operators and their deformations recently found and studied by
  Chalykh, Feigin, Sergeev, and Veselov. We present a unified and
  explicit construction of all these polynomials.

\end{abstract}


\maketitle

\section{Introduction}\label{sec1}
In this paper we discuss the construction of symmetric polynomials
which arise as eigenfunctions of exactly solvable quantum many-body
systems of Calogero-Sutherland type \cite{Cal2,Su2}. In particular, we
demonstrate that a particular construction method found and studied by
us in simple special cases \cite{Lang,HL} can be naturally extended to
the full class of such model. Our results provide a unified approach
to many-variable generalisations of the classical Hermite, Lagueree,
Jacobi and Bessel polynomials.  We also show that it is natural to
generalise our approach to a larger family of polynomials related to a
deformation of the Calogero-Sutherland type systems found and studied
by Chalykh, Feigin, Sergeev, and Veselov \cite{CFV,Sergeev,SV}. This
allows us to derive, for each such polynomial, a family of different
representations labeled by a pair of non-negative integers
$(M,\tilde{M})$. Moreover, one can choose $(M,\tilde{M})$ such that
the representation is simplest, i.e., a linear combination of certain
explicitly given polynomials with the least number of terms. These
results are non-trivial already for the non-deformed polynomials.  To
mention a specific example, we find that a Jack polynomial
\cite{Stanley} has a $(M,\tilde{M})$ representation if there exists a
pair of non-negative integers $(m,\tilde{m})$ such that the Young
diagram corresponding to this Jack polynomial can be covered by the
union of the two rectangular Young diagrams of size $M\times m$ (i.e.,
$M$ rows and $m$ columns) and $\tilde{m}\times\tilde{M}$,
respectively. Moreover, the simplest such representation is obtained
by minimizing $M+\tilde{M}$. For example, the simplest possible
representation of the Jack polynomial corresponding to the Young
diagrams
$$
\mbox{\tiny \Yvcentermath1 $\yng(8)$}\; , \; \mbox{\tiny \Yvcentermath1 $\yng(1,1,1,1,1,1,1)$}\; , \; \mbox{ and }\;  \mbox{\tiny \Yvcentermath1 $\yng(8,7,3,3,3,2,2,2,1)$} 
$$
is for $(M,\tilde{M})=(1,0)$, $(0,1)$, and $(2,3)$, respectively, and
in the first two cases our representation actually consists only of a
single term. The main tool in our construction are certain identities which, to each
(deformed) Calogero-Sutherland system, relates a system of the same
type. In special cases, these identities are well known and related to
a duality of the Jack polynomials (see, e.g., \cite{MacD}), or, more
generally, the algebra homomorphism which maps Jack polynomials to so
called super Jack polynomials (see, e.g., \cite{SV2}). We deduce these
identities in a unified manner, and thus unify, as well as extend,
previously known results.

While we mention these examples in the beginning to give an impression
of the kind of results we obtain, the emphasis in our presentation is
on our construction method. To avoid overloading our formulae we explain
this method in detail for the non-deformed polynomials and the
representations $(M,0)$.  The subsequent extension to the general case is
then remarkably easy.

\subsection{Calogero-Sutherland type models}
As mentioned above, the polynomials studied in this paper play an
important role in physics since they occur in prominent exactly
solvable quantum mechanical models defined by partial differential
operators of Calogero-Sutherland type. These latter operators play a
central role in our construction, and we therefore proceed to explain
their relation to the classical Hermite, Laguerre, Jacobi and Bessel
polynomials. As elaborated below, to each such sequence of
one-variable polynomials $\lbrace p_n: n=0,1,\ldots\rbrace$ there
exists a quantum mechanical model defined by a Schr\"odinger operator
\begin{equation}\label{h} 
  h=-\partial_x^2 + V(x), 
\end{equation}
with a particular potential function $V$ such that its eigenfunctions
are of the form
\begin{equation}\label{psin} 
  \psi_n(x) = \psi_0(x) p_n (z)\, \mbox{ with }\,  z=z(x)
\end{equation}
for particular $C^2$ functions $\psi_0(x)$ and $z(x)$ ($x$ is a real
variable, and we use the notation $\partial_x:=
\frac{\partial}{\partial x}$). Many-variable generalisation of these
polynomials appear in the following generalisation to an arbitrary
number $N$ of particles: each one-particle Schr\"odinger operator
\eqref{h} allows for a many-body generalisation
\begin{equation}\label{HN}
  H_N=\sum_{j=1}^N\left (-\partial_{x_j}^2+V(x_j)\right) +
  \kappa(\kappa-1)\sum_{j<k} W(x_j,x_k)
\end{equation}
with $W$ a particular two-body interaction potential such that $H_N$
has eigenfunctions $\Psi_{\mathbf{n}}$ which are labelled by integer
vectors $\mathbf{n} = (n_1,\ldots,n_N)$ and are of the form
\begin{equation}\label{Psin}
\Psi_{\mathbf{n}}(\mathbf{x}) = \Psi_0(\mathbf{x})
  P_{\mathbf{n}}(\mathbf{z})\; \mbox{ with }\; z_j=z(x_j),
\end{equation}
\begin{equation}\label{Psi0} 
  \Psi_0(\mathbf{x}) = \prod_{j=1}^N \psi_0(x_j) \prod_{j<k}\left(z_k
  - z_j\right)^\kappa
\end{equation}
where the functions $P_{\mathbf{n}}$ are certain symmetric polynomials
in $N$ variables $\mathbf{z}=(z_1,\ldots,z_N)$ generalising the
polynomials $p_n$. To avoid certain technicalities and since it
includes the cases of main physical interests we will assume
$\kappa>0$, even though many of our results actually hold true also
for other values of $\kappa$.  We mention already at this point that
although the symmetry of the polynomials $P_{\mathbf{n}}$ implies that
a complete set of eigenfunctions can be labelled by partitions, i.e.,
sequences $\mathbf{n}=\boldsymbol{\lambda} =
(\lambda_1,\ldots,\lambda_N)$ of non-negative integers $\lambda_j$
such that $\lambda_1\geq\cdots\geq\lambda_N$, it is in our approach
not always natural to make this restriction, as will become evident
below. These operators $H_N$ define exactly solvable quantum-many body
systems of Calogero-Sutherland type \cite{Cal2,Su1,OP2,IMq}; see
Table~\ref{table1} for a list of well-known examples (note that we
have added constants to the potentials $V$ to simplify certain
formulae later on).

\subsection{Deformed Calogero-Sutherland operators}
A further interesting generalisation of the one-particle Schr\"odinger
operators \eqref{h} is the following class of differential operators
in two sets of variables $\mathbf{x}=(x_1,\ldots,x_{N})$ and
${\mathbf{\tilde{x}}}=(\tilde{x}_1,\ldots,\tilde{x}_{\tilde{N}})$,
where $N$ and $\tilde{N}$ are arbitrary non-negative integers:
\begin{multline}\label{D}
H_{N,\tilde{N}} = \sum_{j=1}^{N}\Bigl( -\partial_{x_j}^2 +
  V(x_j)\Bigr) - \sum_{J=1}^{\tilde{N}} \kappa\Bigl(
  -\partial_{\tilde{x}_J}^2 + \tilde{V}(\tilde{x}_J) \Bigr) \\ +
  \kappa(\kappa-1) \sum_{j<k}W(x_j,x_k) +
  \frac{\kappa-1}{\kappa}\sum_{J<K} W(\tilde{x}_j,\tilde{x}_k) \\ +
  (1-\kappa) \sum_{j,K}W(x_j,\tilde{x}_K)
\end{multline}
with a potential function $\tilde{V}$ of the same kind as $V$ but with
different parameters, as specified in Table~\ref{table1}. These
differential operators are natural generalisations of the
Schr\"odinger operator \eqref{HN} in that they also have polynomial
eigenfunctions with corresponding eigenvalues that can be computed
explicitly; see Section~\ref{sec4} for a precise formulation of this
result. In this paper we present an explicit construction of all these
many-variable generalisations of the above mentioned classical
polynomials. We obtain the results by a unified treatment making no
reference to special cases, as explained below.

\begin{table}[b]
  \begin{center}
    \begin{tabular}{|c||c|c|c|}
\hline & & & \\[-1.0ex] & $V(x)$ & $W(x,y)$ & $\tilde{V}(x)$ \\[1.2ex]
\hline

& & & \\[-1.0ex]

I& $\omega^2 x^2-\omega$ & $\frac2{(x-y)^2}$ & $\tilde\omega^2 x^2-\tilde\omega$
\\[1.2ex] 

II & 0 & $\frac1{2\sin^2(\frac{x-y}2)}$ & $(\kappa^2-1)/\kappa^4$ \\[1.2ex]
& & & \\[-1.0ex]

III & $-\frac{c(c+1)}{\cosh^2(x)}+c^2$ &
$\frac1{2\sinh^2(\frac{x-y}2)}-\frac1{2\cosh^2(\frac{x+y}2)}$ &
$-\frac{\tilde{c}(\tilde{c}+1)}{\cosh^2(x)}+\tilde{c}^2$ \\[1.2ex]

& & & \\[-1.0ex]

IV & $\omega^2 x^2 + \frac{a(a-1)}{x^2}$ &
$\frac2{(x-y)^2}+\frac2{(x+y)^2}$ & $\tilde\omega^2 x^2 +
\frac{\tilde{a}(\tilde{a}-1)}{x^2}$ \\ & $- \omega(1+2a)$ & & $-
\tilde{\omega}(1+2\tilde{a})$ \\[1.2ex] 

& & & \\[-1.0ex]

V & $\frac{a(a-1)}{\sin^2 x}-a^2$ & $\frac1{2\sin^2(\frac{x-y}2)}+
\frac1{2\sin^2(\frac{x+y}2)}$ & $\frac{\tilde{a}(\tilde{a}-1)}{\sin^2
x}-\tilde{a}^2$ \\[1.2ex] 

& & & \\[-1.0ex]

VI &$\frac{a(a-1)}{4\sin^2(\frac{x}2)} +
\frac{b(b-1)}{4\cos^2(\frac{x}2)}$ & $\frac1{2\sin^2(\frac{x-y}2)}+
\frac1{2\sin^2(\frac{x+y}2)}$ &
$\frac{\tilde{a}(\tilde{a}-1)}{4\sin^2(\frac{x}2)} +
\frac{\tilde{b}(\tilde{b}-1)}{4\cos^2(\frac{x}2)}$\\ &
$-\frac{(a+b)^2}4$ & & $-\frac{(\tilde{a}+\tilde{b})^2}4$ \\[1.2ex]

& & & \\[-1.0ex]

VII & $\omega^2e^{-2x} $ & $\frac1{2\sinh^2(\frac{x-y}2)}$ &
$\tilde\omega^2e^{-2x}$\\ & $- \omega(1+2c)e^{-x} + c^2$ & & $-
\tilde\omega (1+2\tilde{c})e^{-x} + \tilde{c}^2 $ \\[1.2ex] 

\hline
    \end{tabular}

\bigskip

    \caption{Examples of external- and two-body interaction potentials
    such that the differential operators \eqref{HN} and \eqref{D} have
    polynomial eigenfunctions. The parameters in the last column are:
    $\tilde\omega=-\omega/\kappa$, $\tilde
    c=-(2c+\kappa+1)/(2\kappa)$, $\tilde{a}=-(2a-\kappa-1)/(2\kappa)$
    and $\tilde{b}=-(2b-\kappa-1)/(2\kappa)$.
    \label{table1}
    }
  \end{center}
    
\end{table}

\subsection{Construction method}
To explain the nature of our results we now discuss the standard
Calogero-Sutherland cases $\tilde{N}=0$ in more detail.  Our approach
is based on a set of remarkable identities which provide us with
particular sets of symmetric polynomials on which the action of the
differential operators \eqref{HN} is particularly simple. These latter
polynomials can be defined by the following expansion:
\begin{equation}\label{fn0}
  \frac{\prod_{1\leq j<k\leq N}(1-w_j/w_k)^\kappa}{\prod_{j,k=1}^N
  (1-z_j/w_k)^\kappa} = \sum_{\mathbf{n}\in \mathbb{Z}^N}
  f_{\mathbf{n}}(\mathbf{z}) w_1^{-n_1}w_2^{-n_2}\cdots w_N^{-n_N},
\end{equation}
valid for $|w_N|>|w_{N-1}|>\cdots > |w_1|>\max_k(|z_k|)$. We mention
that these polynomials were first defined in \cite{Lang} by a certain
contour integral. However, using Cauchy's theorem it is easy to verify
that the two definitions are equivalent. We explicitly construct
series representations for the polynomials $P_{\mathbf{n}}$ in
\eqref{Psin} which are of the form $P_{\mathbf{n}}(\mathbf{z}) =
\sum_{\mathbf{m}}
u_{\mathbf{n}}(\mathbf{m})f_{\mathbf{n}}(\mathbf{z})$ and, in
addition, we obtain certain completeness results; see Section 3 for
the precise statements. We will also derive a partial generalisation
of these results to the differential operators $H_{N,\tilde{N}}$; see
Section 4.  We thus generalise, unify, and extend various results
which were known before only in special cases \cite{Lang3,HL}.

We now explain in which sense our approach is unified and give a
precise definition of the classical polynomials we consider: each
such sequence consists of polynomials $p_n$ of order $n=0,1,\ldots$
such that they are common eigenfunctions of a differential operator
\begin{equation}\label{th}
  \tilde{h} = -\psi_0^{-1}h\psi_0 = \alpha(z)\partial_z^2 +
  \beta(z)\partial_z
\end{equation}
with
\begin{equation}\label{ab}
  \alpha(z) = \alpha_2z^2 + \alpha_1z + \alpha_0\quad
  \textrm{and}\quad \beta(z) = \beta_1z + \beta_0
\end{equation}
for some real coefficients $\alpha_j$ and $\beta_j$; see
Table~\ref{table2} for the polynomials corresponding to our examples
in Table \ref{table1}. It is interesting to note that the
Schr\"odinger operator $h$ in \eqref{h} with the eigenfunctions
\eqref{psin} can be fully characterised by these polynomials $\alpha$
and $\beta$: it is straightforward to verify that
\begin{equation}\label{V1} 
  V(x)=v(z(x)),\quad v(z) =
  \frac{(2\beta(z)-\alpha'(z))(2\beta(z)-3\alpha'(z))}{16\alpha(z)} - \frac14\alpha''(z)
  + \frac12\beta'(z)
\end{equation}
(the prime here and in the following indicates differentiation) with
$z(x)$ a solution of the differential equation
\begin{equation}\label{R2}
  z'(x)^2 = \alpha(z(x))
\end{equation}
and
\begin{equation}\label{psi01}
  \psi_0(x)=e^{-w(z(x))}
\end{equation}
with $w(z)$ any solution of the differential equation
\begin{equation*}
	w'(z)=\frac{\alpha(z)'-2\beta(z)}{4\alpha(z)}.
\end{equation*}
Moreover, the potential functions $W$, ensuring that the differential
operators $H_N$ in \eqref{HN} and $H_{N,\tilde{N}}$ in \eqref{D} have
polynomial eigenfunctions, are given by
\begin{equation}\label{W} 
  W(x_1,x_2) = \frac{\alpha(z_1)+\alpha(z_2)}{(z_1-z_2)^2}-\alpha_2\;
  \mbox{ with $z_j=z(x_j)$}.
\end{equation}
The constant $\alpha_2$ is subtracted to simplify formulae later on,
and $\tilde{V}$ is given by a formula as in \eqref{V1} but with $\beta(z)$
replaced by
\begin{equation*}
  \tilde\beta(z)= [(1+\kappa)\alpha'(z)-\beta(z)]/\kappa.
\end{equation*}
It is interesting to note that the exact eigenvalues of the
differential operators $H_N$ in \eqref{HN} and $H_{N,\tilde{N}}$ in
\eqref{D} are determined by the leading coefficients of these
polynomials $\alpha(z)$ and $\beta(z)$. For example, the exact eigenvalues
of $H_N$ corresponding to the eigenfunctions \eqref{Psin} are
\begin{equation}\label{En} 
  E_{\mathbf{n}} = E_0 - \sum_{j=1}^N\left( \alpha_2 n_j(n_j-1)+(
  2\kappa\alpha_2(N-j)+ \beta_1 ) n_j \right)
\end{equation}
with
\begin{equation}\label{E0}
  E_0 = -\frac{\alpha_2\kappa^2}3 N(N^2-1) - \frac{\kappa(\beta_1 - (1
  + \kappa)\alpha_2)}2 N(N-1).
\end{equation}
This result, well-known in many special cases (see, e.g., \cite{GGR}),
will naturally emerge from our construction in Section~\ref{sec3}.  We
are now in a position to state more precisely in which sense our
approach is unified: in our construction information concerning the
different cases will enter only via the two polynomials $\alpha$ and
$\beta$.  We also stress that the special cases listed in Table
\ref{table2} are only intended as examples, and that our approach is
for arbitrary polynomials $\alpha$ and $\beta$ of the form \eqref{ab}.

\begin{table}[htbp]
  \begin{center}
    \begin{tabular}{|c||c|c|c|l|l|}
\hline 
& & & & & \\[-1.0ex]
& $\psi_0(x)$ & $z(x)$ & $p_n(z)$ & $\alpha(z)$ & $\beta(z)$ \\[1.2ex] \hline 

& & & & & \\[-1.0ex]

I& $e^{-\omega x^2/2}$ & $x$ & $H_n(\sqrt{\omega}z)$ & $1$ & $-2\omega z$ \\
& & & (Hermite) & & \\[1.2ex]

II & $1$ & $e^{ix}$ & $z^n$ & $-z^2$ & $-z$ \\
& & & & & \\[1.2ex]

III & $\cosh^{-c}(x)$ & $i\sinh(x)$ &
$C_n^{(-c)}(z)$ & $-1+z^2$ & $(1-2c)z$ \\
& & & (Gegenbauer) & & \\[1.2ex]

IV & $e^{-\omega x^2/2} x^{a}$ & $x^2$ &
$L_n^{(a-\frac12)}(\omega z)$ & $4 z$ & $2 + 4 a -4\omega z$ \\
& & & (Laguerre) & & \\[1.2ex] 

V &
$\sin^a(x)$ & $\cos(x)$ & $C_n^{(a)}(z)$ & $1-z^2$ & $-(1+2a)z$ \\
& & & (Gegenbauer) & & \\[1.2ex] 

VI &
$\sin^{a}(\frac{x}2)\cos^{b}(\frac{x}2)$ & $\cos(x)$ &
$P_n^{(a-\frac12,b-\frac12)}(z)$ & $1-z^2$ & $\quad b-a- $ \\
& & & (Jacobi)& & $(1+a+b)z$ \\[1.2ex]

VII & $\exp{(-\omega e^{-x} - c x)}$&
$e^x$ & $y_n(z,1-2c,2\omega)$ & $z^2$ & $2\omega + (1-2 c)z$ \\ 
& & & (gen.\ Bessel) & & \\[1.2ex] 

\hline
    \end{tabular}

\bigskip

    \caption{Exact solutions of the one-body Schr\"odinger equation
    associated with classical polynomials, as described in \eqref{h}
    and \eqref{psin}. Given are also the associated polynomials
    $\alpha$ and $\beta$; see \eqref{th}--\eqref{psi01}. More details
    about the polynomials $p_n$ can be found in \cite{Abr} (cases~I
    and III--VI) and in \cite{Gross} (case~VII).
    \label{table2}}
  \end{center}    
\end{table}

\subsection{Special cases}
To put this general scheme into perspective we now discuss its
well-known special cases summarised in Tables~\ref{table1} and
\ref{table2}. The case $\alpha = 1$ and $\beta(z) = -2\omega z$
corresponds to the exactly solvable many-body generalisation of the
quantum harmonic oscillator introduced by Calogero \cite{Cal2} (case~I
in Tables~\ref{table1} and \ref{table2}), while the Sutherland model
\cite{Su1,Su2}, generalising the quantum model of a free particle on a
circle, is obtained by setting $\alpha(z) = -z^2$ and $\beta(z) = -z$
(case~II). These models correspond to the Hermite polynomials and the
ordinary monomials $p_n(z)=z^n$, respectively. The many-body models
associated with the Legendre, Gegenbauer and Jacobi polynomials
(cases~IV, V and VI) are identical with Olshanetsky and Perelomov's
$B_N$- and $BC_N$-variants of the Calogero- and Sutherland models; see
\cite{OP2} and references therein. An exact solution to the many-body
generalisation of the Morse potential (case~VII), corresponding to the
generalised Bessel polynomials \cite{KF}, was first given by
Inozemtsev and Meshcheryakov \cite{IMq}. A further study of this case
can be found in \cite{Hal3}. There exist various other interesting
special cases which we have not explored in detail, but they can all
be transformed to the cases mentioned above by a rescaling and a
translation of the variable $z$. The deformations of the
Calogero-Sutherland models defined by $H_{N,\tilde{N}}$ in \eqref{D}
were found and explored by Chalykh, Feigin, Sergeev, and Veselov; see,
e.g., \cite{CFV,Sergeev,SV,SV2}. We will refer to these differential
operators $H_{N,\tilde{N}}$ as deformed Calogero-Sutherland operators.

\subsection{Related previous work}
We now discuss the relation of our results to previous work in the
literature. The many-variable polynomials corresponding to the
eigenfunctions of the Calogero-Sutherland type models in cases~I--VI
in Table~\ref{table1} have been extensively studied in the mathematics
literature; see, e.g., \cite{DunklXu,MacD} and references therein.  We
mention, in particular, Heckman and Opdam's root system generalisation
of the Jacobi polynomials \cite{HeckOp} and the work of Baker and
Forrester \cite{BF}, van Diejen \cite{Die}, and Lassalle
\cite{Las,Las2,Las3}, as well as of Macdonald \cite{MacD2}, on
many-variable generalisations of the classical orthogonal Hermite,
Laguerre and Jacobi polynomials. A particularly well-studied case are
the so-called Jack polynomials corresponding to the Sutherland model
(case~II).  Explicit formulae for the Jack polynomials were recently
obtained by Lassalle and Schlosser \cite{LS} (see also \cite{Las4}) by
inverting a so-called Pieri formula. For very particular values of the
integer vector $\mathbf{n} = (n_1,\ldots,n_N)$ or a low number of
variables explicit series expansions of the Jack polynomials were
obtained by Stanley \cite{Stanley}.  In addition, there exist
expansions of the Jack, as well as certain other many-variable
classical orthogonal polynomials, which are of a combinatorial nature
\cite{DLM,DLM2,KS,MacD}.
These combinatorial results also include a formula resulting from the
use of Sutherland's original solution algorithm \cite{Su2}. We also
mention operator solutions of the Calogero- and Sutherland models
obtained in \cite{BHV,Kak,Kak2,UW,UW2} and \cite{LV}, respectively, as
well as integral representations of the Jack polynomials; see
\cite{AMOS,MY,OO} and references therein. Our results in this paper
provide an explicit construction of the many-variable polynomials
$P_{\mathbf{n}}$. These results were recently announced in
\cite{Hal1}, and they generalise those previously obtained by us for
the Sutherland- \cite{Lang} and the Calogero models \cite{HL}. We
mention that our unified treatment of all the cases in
Table~\ref{table1} is different from the one based on root systems
(see, e.g., \cite{OP2}), and that it has been previously used by
Gomez-Ullate et.al.\ \cite{GGR} to obtain the energy eigenvalues in
\eqref{En} and \eqref{E0} by purely algebraic means. We also mention
that our point of view is closely related to the theory of
quasi-exactly solvable Schr\"odinger operators; see, e.g.,
\cite{BTO,GKO}. Moreover, the identities which are the key to our
results are stated in Corollaries~\ref{cor2} and \ref{cor4} below, and
they were known before only in special cases: an important special
case of these identities for the Sutherland model (case~II) is a
consequence of a well-known result on Jack polynomials which, to our
knowledge, is due to Stanley (see Proposition~2.1 in \cite{Stanley}),
and a generalisation of the latter to the deformed case and other
non-deformed cases can be found in \cite{SV2} and
\cite{MacD,Gaudin,Ser}, respectively.  These identities relate
Schr\"odinger- or deformed Calogero-Sutherland operators with
different parameter values and are natural quantum analogs of the
B\"acklund transformations for the classical Calogero-Moser models
first found by Woijekowski \cite{W}, as discussed by Kutznetsov and
Sklyanin \cite{KuzS}. We obtain these identities as specialisations of
a particular identity stated in Proposition~\ref{prop1} which has the
natural physical interpretation of giving the groundstate of a
generalisation of the Schr\"odinger operator \eqref{HN} where the
particles are allowed to have different masses $m_j$ and with specific
mass dependent external potentials $V_{m_j}(x_j)$. This is a powerful
result of independent interest, and to our knowledge it was previously
known only in the special cases~I, II and IV in Table~\ref{table1}
\cite{Sen,F,HL}. We finally mention an integral representation of the
Jack polynomials recently obtained by Kuznetsov et.al.\ \cite{KMS}
using a separation-of-variables approach which is also based on the
identity in Corollary~\ref{cor2}.

\subsection{Plan of the paper}
In Section~\ref{sec2} we derive and discuss the identities which are
the key to our solution method.  Section~\ref{sec3} contains our
explicit construction of series representation of all many-variable
polynomials $P_{\mathbf{n}}$ determining the eigenfunctions
\eqref{Psin} of the Schr\"odinger operators \eqref{HN}. In
Section~\ref{sec4} we extend our construction of polynomial
eigenfunctions to the deformed Calogero-Sutherland operators
\eqref{D}. We conclude with a few remarks in Section~\ref{sec5}.

\subsection{Notation} We denote by $\mathbb{Z}$, $\mathbb{N}$,
$\mathbb{N}_0$ and $\mathbb{R}$ the sets of all integers, positive
integers, non-negative integers and real numbers, respectively. We
shall say that an integer vector $\mathbf{n} = (n_1,n_2,\ldots)$ is of
length $N$, denoted $\ell(\mathbf{n}) = N$, if $N$ is the smallest
non-negative integer such that the parts $n_j$ of $\mathbf{n}$ are
zero for $j>N$. The symbols $\mathbf{n}$ and $\mathbf{m}$ will in most
instances denote integer vectors, except in Section~\ref{sec2} where
the symbols $m_j$ are used for real `mass parameters' rather than
for parts of an integer vector $\mathbf{m}$. We will use the symbols
$\boldsymbol{\lambda}$, $\boldsymbol{\mu}$ and $\boldsymbol{\nu}$ to
emphasise that a particular integer vector is a partition.  We will
also use the notation
\begin{equation*}
\mathbf{x}^{\mathbf{s}} = x_1^{s_1}\cdots x_N^{s_N}
\end{equation*}
where $\mathbf{x} = (x_1,\ldots,x_N)$ and $\mathbf{s} =
(s_1,\ldots,s_N)$. Moreover, for $\mathbf{m}\in\mathbb{Z}^M$ and
$\mathbf{n}\in\mathbb{Z}^N$ we will write $(\mathbf{m},\mathbf{n})$
short for $(m_1,\ldots,m_M,n_1,\ldots,n_N)$.

\section{Identities}\label{sec2} 
In this section we present and prove a particular identity associated
with the one-particle Schr\"odinger operator \eqref{h}. The identities
underlying our construction of eigenfunctions of the differential
operators in \eqref{HN} and \eqref{D} are stated in
Section~\ref{sec22} and obtained as special cases of this more general
results. Throughout the section, we assume the polynomials $\alpha$
and $\beta$ in \eqref{ab} fixed, and $z(x)$ and $W(x_1,x_2)$ are as in
\eqref{R2} and \eqref{W}.

\subsection{Source identity}\label{sec21}
The following identity can be interpreted as providing the exact
groundstate of a generalisation of the Schr\"odinger operator
\eqref{HN} where the particles can have different masses $m_j$.
However, we will allow these parameters $m_j$ also to be negative, and
we will use this identity as a source from which we obtain various
other identities as special cases. To state this result we find it
convenient to denote this generalised Schr\"odinger operators as $\mathcal{H}$,
the particle coordinates as $X_j$, and the particle number as
$\mathcal{N}$.

\begin{prop}\label{prop1} Let
\begin{equation}\label{cH}
  \mathcal{H} = \sum_{j=1}^{\mathcal{N}} \frac1{m_j}
  \left(-\partial_{X_j}^2 + V_{m_j}(X_j) \right) + \sum_{j<k}
  \frac{\kappa}{2}(\kappa m_j m_k-1)(m_j+m_k) W(X_j,X_k)
\end{equation}
with $m_j$ arbitrary real and non-zero parameters and
\begin{equation}\label{Vm}
  V_m(X)= v_m(z(X))
\end{equation}
with
\begin{equation*}
  v_m(z)=
  \frac{(2\beta_m(z)-\alpha'(z))(2\beta_m(z)-3\alpha'(z))}{16\alpha(z)} - \frac14
  \alpha''(z) + \frac12 \beta'_m(z),
\end{equation*}
where the prime indicates differentiation with respect to the argument
$z$ and
\begin{equation}\label{betam}
  \beta_m(z)= m \beta(z)+\frac12(1-m)(1-\kappa m) \alpha'(z).
\end{equation}
Furthermore, let
\begin{equation*}
  \Phi_0(\mathbf{X}) = \prod_{j=1}^{\mathcal{N}} \psi_{0,m_j}(X_j)
  \prod_{j<k}(Z_k - Z_j)^{\kappa m_j m_k}
\end{equation*}
with $Z_j = z(X_j)$ and 
\begin{equation}\label{wpm} 
 \psi_{0,m}(X)=e^{-w_m(z(X))}
\end{equation}
with $w_m(z)$ any solution of the differential equation
\begin{equation*}
	w'_m(z)= \frac{\alpha'(z)-2\beta_m(z)}{4\alpha(z)}.
\end{equation*}
Then 
\begin{equation}\label{MID} 
  \left( \mathcal{H} -\mathcal{E}_0 \right) \Phi_0 = 0
\end{equation}
with the constant
\begin{equation}\label{cE0} 
\mathcal{E}_0 = -\frac{g^2\alpha_2}{3}(|\mathbf{m}|^3 -
|\mathbf{m}^3|) - \frac{\kappa(\beta_1 -
  (1+\kappa)\alpha_2)}2(|\mathbf{m}|^2 - |\mathbf{m}^2|),
\end{equation}
where
\begin{equation}\label{vmn} 
  |\mathbf{m}^n| := \sum_{j=1}^{\mathcal{N}} m_j^n\; \mbox{ for
   $n=1,2,3$}.
\end{equation}
\end{prop}

The $m$-dependence of the external potentials and one-particle
groundstate eigenfunctions for our examples is given in
Table~\ref{table3} with $m$-dependent parameters defined in the table
captions.

\begin{remark}\label{betaremark}
Note that $\beta_1$ and $\beta_0$ always refers to the coefficients of
the polynomial $\beta(z)$ as defined in \eqref{ab}, and are not to be
confused with $\beta_m(z)$ for $m=1$ and $m=0$. There should be no
danger of confusion since we always write $\beta(z)$ for $\beta_m(z)$
if $m=1$ and assume $m\neq 0$.
\end{remark} 

\begin{table}[htbp]
  \begin{center}
    \begin{tabular}{|c|c|c|}
\hline & & \\[-1.0ex]

& $V_m(x)$ & $\psi_{0,m}(x)$\\[1.2ex] \hline 
& & \\[-1.0ex]

I& $\omega_m^2 x^2-\omega_m$ & $e^{-\omega_m x^2/2}$\\[1.2ex]

II & $-\frac14 \kappa^2 m^2(m-1)^2$ & $e^{i\kappa m(m-1)x/2}$  \\[1.2ex]

III & $-\frac{c_m(c_m+1)}{\cosh^2(x)} + c_m^2$ & $\cosh^{-c_m}(x)$ 
\\[1.2ex]

IV & $\omega_m^2 x^2 +
\frac{a_m(a_m-1)}{x^2}- \omega_m(1+2a_m)$ & $e^{-\omega_m x^2/2} x^{a_m}$
\\[1.2ex] 

V & $\frac{a_m(a_m-1)}{\sin^2 x} -a_m^2$ & $\sin^{a_m}(x)$ 
\\[1.2ex]

VI &$\frac{a_m(a_m-1)}{4\sin^2(\frac{x}2)} +
\frac{b_m(b_m-1)}{4\cos^2(\frac{x}2)} - \frac14(a_m+b_m)^2$ &
$\sin^{a_m}(\frac{x}2)\cos^{b_m}(\frac{x}2)$ \\[1.2ex]

VII & $\omega_m^2e^{-2x} - \omega_m(1+2c_m)e^{-x} + c_m^2$ & $e^{-\omega_m
  e^{-x} - c_m x}$ \\[1.2ex] 

\hline
    \end{tabular}

\bigskip

\caption{`Mass' dependence of the external potentials and one-particle
groundstates in our examples, according to Proposition~\ref{prop1}.
The $m$-dependence of the parameters is as follows, $\omega_m = m\omega$,
$c_m = mc-\mbox{$\frac12$} \kappa m(m-1)$, $a_m = ma +
\mbox{$\frac12$} \kappa m(m-1)$, and $b_m = mb + \mbox{$\frac12$}
\kappa m(m-1)$.
\label{table3}}
  \end{center}
\end{table}

\begin{proof}[Proof of Proposition~\ref{prop1}]
We will show by straightforward computations that
\begin{equation}\label{ID}
  \mathcal{H} = \sum_{j=1}^{\mathcal{N}} \frac1{m_j} Q_j^+ Q_j^-
  +\mathcal{E}_0,
\end{equation}
where
\begin{equation*}
Q_j^\pm = \mp \partial_{X_j} + \mathcal{V}_j,\quad
    \mathcal{V}_j(\mathbf{x}) = \frac{1}{\Phi_0(\mathbf{x})}
    \partial_{X_j} \Phi_0(\mathbf{x}).
\end{equation*} 
Since $Q_j^-\Phi_0=0$ for all $j$, this will prove the identity in
\eqref{MID}.

To prove \eqref{ID} we compute
\begin{equation*}
  \mathcal{V}_j(\mathbf{x}) = -w'_{m_j}(z(X_j))z'(X_j) + \sum_{k\neq
  j} \kappa m_j m_k \frac{z'(X_j)}{z(X_j)-z(X_k)},
\end{equation*}
and thus 
\begin{equation*}
  \sum_{j=1}^{\mathcal{N}} \frac1{m_j} Q_j^+ Q_j^- = -
  \sum_{j=1}^{\mathcal{N}} \frac1{m_j} \partial_{X_j}^2 +
  \mathcal{W}_1+\mathcal{W}_2+ \mathcal{W}_3
\end{equation*}
with
\begin{equation*}
  \mathcal{W}_1 = \sum_{j=1}^{\mathcal{N}} \frac1{m_j}\left(
  w'_{m_j}(Z_j)^2\alpha(Z_j) - w_{m_j}''(Z_j)\alpha(Z_j) - \frac12
  w'_{m_j}(Z_j) \alpha'(Z_j) \right)
\end{equation*}
the one-body terms,
\begin{equation*}
  \mathcal{W}_2 = \sum_{j=1}^{\mathcal{N}} \sum_{k\neq j} \kappa m_k
  \left( (\kappa m_jm_k-1)\frac{\alpha(Z_j)}{(Z_j-Z_k)^2} +
  \frac{\frac12\alpha'(Z_j)-2w_{m_j}'(Z_j)\alpha(Z_j)}{Z_j-Z_k}
  \right)
\end{equation*}
the two-body terms, and
\begin{equation*}
  \mathcal{W}_3 = \sum_{j=1}^{\mathcal{N}} \sum_{k,\ell\neq j, k\neq
  \ell} \kappa^2 m_jm_km_\ell \frac{\alpha(Z_j)}{(Z_j-Z_k)
  (Z_j-Z_\ell)}
\end{equation*}
the three-body terms; we inserted the relations
$z'(X_j)^2=\alpha(Z_j)$ and $z''(X_j)=\frac12\alpha'(Z_j)$ and used
the short hand notation $z(X_j)=Z_j$.

Symmetrising the sum defining $\mathcal{W}_3$ and inserting
$\alpha(Z_j) = \sum_{p=0}^2\alpha_p Z_j^p$ we can write
\begin{equation*}
  \mathcal{W}_3 = \sum_{j <k <\ell} 2 \kappa^2
  m_jm_km_\ell\sum_{p=0}^2 \alpha_p \frac{Z_j^p(Z_k-Z_\ell) -
  Z_k^p(Z_j-Z_\ell) + Z_\ell^p(Z_j-Z_k)}{(Z_j-Z_k)
  (Z_j-Z_\ell)(Z_k-Z_\ell)},
\end{equation*}
and observing that the fraction on the r.h.s.\ is identical with $0$ and
$1$ for $p=0,1$ and $p=2$, respectively, we find that $\mathcal{W}_3$ is
equal to the constant
\begin{equation*}
  \mathcal{W}_3 = \sum_{j <k <\ell} 2\kappa^2 m_jm_km_\ell \alpha_2.
\end{equation*}

Symmetrising the sum defining $\mathcal{W}_2$ and inserting
$\frac12\alpha'(Z_j)-2w'_{m_j}(Z_j)\alpha(Z_j)=\beta_{m_j}(Z_j)$,
following from the second equation in \eqref{wpm}, we obtain
\begin{equation*}
  \mathcal{W}_{2} = \sum_{j<k} \kappa \left( (\kappa m_jm_k-1)
  \frac{m_k \alpha(Z_j) + m_j\alpha(Z_k)}{(Z_j-Z_k)^2} +
  \frac{m_k\beta_{m_j}(Z_j)-m_j\beta_{m_k}(Z_k)}{Z_j-Z_k}\right).
\end{equation*}
We now decompose $\mathcal{W}_{2}$ into two parts as follows:
\begin{equation*}
  \mathcal{W}_{2} = \mathcal{W}_{2,0} + \mathcal{W}_{2,1},
\end{equation*}
where
\begin{equation*}
  \mathcal{W}_{2,0} = \sum_{j < k} \frac{\kappa}2(\kappa
  m_jm_k-1)(m_j+m_k)\frac{\alpha(Z_j) + \alpha(Z_k)}{(Z_j-Z_k)^2}
\end{equation*}
and
\begin{multline*}
  \mathcal{W}_{2,1} = \sum_{j<k} \kappa \Biggl( \frac12 (\kappa
  m_jm_k-1)(m_k-m_j)\frac{\alpha(Z_j) - \alpha(Z_k)}{(Z_j-Z_k)^2}\\ +
  \frac{m_k\beta_{m_j}(Z_j)-m_j\beta_{m_k}(Z_k)}{Z_j-Z_k} \Biggr).
\end{multline*}
Recalling \eqref{W} we find that the symmetric part gives us the
two-body terms we want, up to a constant,
\begin{equation*}
  \mathcal{W}_{2,0} = \sum_{j<k} \frac{\kappa}2(\kappa
  m_jm_k-1)(m_j+m_k)\left( W(x_j,x_k) + \alpha_2 \right).
\end{equation*}
Inserting \eqref{betam} and \eqref{ab} a straightforward but somewhat
tedious computation shows that the terms in the antisymmetric term add
up to a constant,
\begin{equation*}
  \mathcal{W}_{2,1} = \sum_{j<k}\kappa\left( \frac{\alpha_2}2 (\kappa
  m_jm_k+1) (m_j+m_k) + (\beta_1 - \alpha_2(1+\kappa))m_jm_k \right) .
\end{equation*}
Inserting the second equation in \eqref{wpm} and \eqref{betam} a
simple computation shows that the one-body terms are identical with
\begin{equation*}
\mathcal{W}_1 = \sum_j \frac1{m_j} V_{m_j}(X_j)
\end{equation*}
with $V_{m}$ defined in \eqref{Vm}. Collecting all terms we obtain the
identity in \eqref{ID} with the constant
\begin{multline*}
  \mathcal{E}_0 = -2\kappa^2\alpha_2\sum_{j<k<\ell} m_j m_k m_\ell -
  \kappa^2\alpha_2 \sum_{j<k} m_jm_k(m_j+m_k)\\ - \kappa(\beta_1 -
  \alpha_2(1+\kappa))\sum_{j<k} m_jm_k .
\end{multline*}
Using the notation in \eqref{vmn} we find by straightforward
computations that this constant is identical with the one given in
\eqref{cE0}.
\end{proof} 

The physical interpretation of $\mathcal{H}$ in \eqref{cH} as
Schr\"odinger operator of a quantum many-body system requires that it defines a
self-adjoint Hilbert space operator bounded from below. This is the
case under certain obvious restrictions on parameters. We now discuss
this Hilbert space structure for the different cases listed in
Table~\ref{table1}, but our discussion will be brief since this aspect
does not play any role for our construction in the following sections.

In cases~I, III and VII the relevant Hilbert space is
$L^2(\mathbb{R}^{\mathcal{N}})$, in case~IV we have instead
$L^2(\mathbb{R}_+^{\mathcal{N}})$, while
$L^2(\lbrack-\pi,\pi\rbrack^{\mathcal{N}})$ is associated to case~II,
and $L^2(\lbrack 0,\pi\rbrack^{\mathcal{N}})$ to cases~V and VI (the
weight function is in all these cases constant and equal to 1). It is
obvious in all these cases that $Q_j^+$ in the proof above is the
Hilbert space adjoint of $Q_j^-$ (on suitable domains), and thus, if
all $m_j$ are positive, that $\mathcal{H}$ in (\ref{ID}) defines a
unique self-adjoint operator via the Friedrichs extension (see, e.g.,
\cite{RS2}) with $\Phi_0$ as groundstate provided that $\Phi_0$ is
square integrable.  In particular, the $H_N$ in (\ref{HN}) define,
under obvious restrictions on parameters, self-adjoint operators
bounded from below. However, this is not the case for the deformed
operators (\ref{D}).

\subsection{Important special cases}\label{sec22} 
We proceed to discuss the special cases of Proposition~\ref{prop1}
which underlies our construction of eigenfunctions of the differential
operators \eqref{HN} and \eqref{D} in Sections~\ref{sec3} and
\ref{sec4}, respectively. Setting $\mathcal{N} = N$, $m_j = 1$ and
$X_j = x_j$ for $j = 1,\ldots,N$ we obtain as a first special case the
following:

\begin{cor}
The function $\Psi_0$, as defined by \eqref{Psi0} and \eqref{psi01},
is an eigenfunction of the Schr\"odinger operator \eqref{HN} with
corresponding eigenvalue $E_0$.
\end{cor}

The remarkable identity underlying our solution method in
Section~\ref{sec3} is obtained from Proposition~\ref{prop1} by
choosing $\mathcal{N}=2N$, $m_j=1$, $X_j=x_j$, $m_{N+j}=-1$, and
$X_{N+j}=y_j$ for $j=1,\ldots,N$. We observe that $\mathcal{H}$ in
\eqref{cH} then becomes a difference of two Schr\"odinger operators
\eqref{HN}. Denoting $\Phi_0$ as $F$ and $\mathcal{E}_0$ as $C_{N}$
we obtain the following:

\begin{cor}\label{cor2}
Let $H_N(\mathbf{x})=H_N$ be the operator in \eqref{HN},
\begin{equation}\label{Hminus}
  H^{(-)}_{N}(\mathbf{y}) = \sum_{j=1}^N\Big(-\partial_{y_j}^2 +
 V_{-1}(y_j)\Big) + \kappa(\kappa - 1)\sum_{j<k}W(y_j,y_k)
\end{equation}
and 
\begin{equation*}
\begin{split}
  F(\mathbf{x},\mathbf{y}) = \prod_{j=1}^N\psi_{0}(x_j)
  \psi_{0,-1}(y_j) \frac{\prod_{j<k}(z_k-z_j)^{\kappa}
  (w_k-w_j)^{\kappa} }{\prod_{j,k}(w_k-z_j)^{\kappa}}
\end{split}
\end{equation*} 
with $z_j=z(x_j)$, $w_j=z(y_j)$, and $V_{-1}(x)$ and $\psi_{0,-1}(x)$
as in \eqref{Vm} and \eqref{wpm} for $m=-1$. Then
\begin{equation}\label{Id} 
  \left( H^{\phantom\dag}_{N}(\mathbf{x}) - H^{(-)}_{N}(\mathbf{y})
  - C_{N}\right) F(\mathbf{x},\mathbf{y}) = 0
\end{equation}
with the constant 
\begin{equation*}
  C_N = \kappa(\beta_1 - \alpha_2(1+\kappa) ) N .
\end{equation*}
\end{cor}

It is important to note that also $H^{(-)}_N$ is a Schr\"odinger
operator of the same kind as $H_N$, only the coupling parameters are
different. For the convenience of the reader we give the modified
parameters for our examples I and III--VII,
\begin{equation*}
  \omega_{-1} =-\omega,\quad c_{-1}= -\kappa-c, \quad a_{-1} =
  \kappa-a,\quad b_{-1}= \kappa-b.
\end{equation*}

The key property that makes the special case of Proposition
\ref{prop1} stated in Corollary~\ref{cor2} interesting as a tool for
constructing the eigenfunctions of a Schr\"odinger operator \eqref{HN}
is that the variables $\mathbf{X}$ of $\mathcal{H}$ in \eqref{cH} are
divided in two groups $\mathbf{x}$ and $\mathbf{y}$ such that the
interaction terms involving variables belonging to different groups
all vanish. It is interesting to note that there are other cases where
this happens. For example, the number of particles with parameters
$m_j=1$ and $m_j=-1$ need not be the same. More generally, if we
divide the variables $\mathbf{X}$ into two groups where the `mass
parameters' in the first group are $m_j=1$ or $-1/\kappa$ and in the
second group $m_k=-1$ or $1/\kappa$, then all interaction terms
between particles in different groups vanish (since $\mbox{$\frac12$}
\kappa (m_j+m_k)(\kappa m_j m_k -1)=0$ in all these cases). We then
obtain an identity as in \eqref{Id} but now involving two deformed
Calogero-Sutherland operators \eqref{D} which can have, in general,
different particle numbers. Such an identity involves four different
kinds of particles, and the number of particles of each kind can be
arbitrary. These identities will allow us to construct eigenfunctions
of the deformed Calogero-Sutherland operators \eqref{D} in
Section~\ref{sec4}.  The most general such identity corresponds to the
following special case of Proposition~\ref{prop1}:
$\mathcal{N}=N+\tilde{N}+M+\tilde{M}$ with non-negative integers
$N,\tilde{N},M,\tilde{M}$ such that $\mathcal{N}$ is non-zero, $m_j=1$
and $X_j=x_j$ for $j=1,2,\ldots,N$, $m_{N+J}=-1/\kappa$ and
$X_{N+J}=\tilde{x}_J$ for $J=1,2,\ldots,\tilde{N}$, $m_{N+\tilde{N} +
k}=-1$ and $X_{N+\tilde{N} + k} = y_k$ for $k=1,2,\ldots,M$, and
$m_{N+ \tilde{N}+M + K}= 1/\kappa$ and $X_{N +\tilde{N} + M +
K}=\tilde{y}_K$ for $K=1,2,\ldots,\tilde{M}$.

\begin{cor}\label{cor4} 
Let $H_{N,\tilde{N}}(\mathbf{x},{\mathbf{\tilde{x}}})=H_{N,\tilde{N}}$
be the operator in \eqref{D} with $\tilde{V}=V_{-1/\kappa}$,
\begin{multline*}
 H^{(-)}_{M,\tilde{M}}(\mathbf{y},{\mathbf{\tilde{y}}}) =
 \sum_{k=1}^{M} \Big(-\partial_{y_k}^2 + V_{-1}(y_k)\Big) -
 \sum_{K=1}^{\tilde{M}} \kappa\Big(-\partial_{\tilde{y}_K}^2 +
 V_{1/\kappa}(\tilde{y}_K)\Big) \\ + \kappa(\kappa-1)
 \sum_{j<k}W(y_j,y_k) + \frac{\kappa-1}{\kappa}\sum_{J<K}
 W(\tilde{y}_j,\tilde{y}_k) \\ + (1-\kappa)
 \sum_{j,K}W(y_j,\tilde{y}_K)
\end{multline*}
where $M$ and $\tilde{M}$ are arbitrary non-negative integers, and
\begin{multline*}
F_{N,\tilde{N},M,\tilde{M}}(\mathbf{x},\mathbf{\tilde{x}}, \mathbf{y},
  \mathbf{\tilde{y}}) = \prod_{j=1}^{N}\psi_{0}(x_j)
  \prod_{J=1}^{\tilde{N}} \psi_{0,-1/\kappa}(\tilde{x}_J)
  \prod_{k=1}^{M} \psi_{0,-1}(y_k) \prod_{K=1}^{\tilde{M}}
  \psi_{0,1/\kappa}(\tilde{y}_K) \\ \quad \times
  \frac{\prod_{j<k}(z_k-z_j)^{\kappa} \prod_{J<K}(
  \tilde{z}_K-\tilde{z}_J)^{1/\kappa}}{\prod_{j,J}(\tilde{z}_J-z_j)}
  \\ \quad\times \frac{ \prod_{j<k}(w_k-w_j)^{\kappa} \prod_{J<K}(
  \tilde{w}_K-\tilde{w}_J)^{1/\kappa}}{\prod_{j,J}(\tilde{w}_J-w_j)}\\
  \quad \times \frac{\prod_{j,K}(\tilde{w}_K-z_j)
  \prod_{J,k}(w_k-\tilde{z}_J)}{\prod_{j,k}(w_k-z_j)^\kappa
  \prod_{J,K}(\tilde{w}_K-\tilde{z}_J)^{1/\kappa}}
\end{multline*}
with $z_j=z(x_j)$, $w_j=z(y_j)$, $\tilde{z}_J=z(\tilde{x}_J)$,
$\tilde{w}_J=z(\tilde{y}_J)$, and $V_{m}(x)$ and $\psi_{0,m}(x)$ as
in \eqref{Vm} and \eqref{wpm}, respectively, for $m=-1$ and $m=\pm
1/\kappa$. Then
\begin{equation*}
  \left( H^{\phantom\dag}_{N,\tilde{N}}(\mathbf{x},
  \mathbf{\tilde{x}}) - H^{(-)}_{M,\tilde{M}}( \mathbf{y},
  {\mathbf{\tilde{y}}} ) - C_{N,\tilde{N},M,\tilde{M}} \right)
  F_{N,\tilde{N},M,\tilde{M}}(\mathbf{x}, {\mathbf{\tilde{x}}},
  \mathbf{y}, {\mathbf{\tilde{y}}}) = 0
\end{equation*}
with the constant 
\begin{multline*}
  C_{N,\tilde{N},M,\tilde{M}}= -\frac{\kappa^2\alpha_2 }3\left(
  \left(N_- - \tilde{N}_-/\kappa \right)^3 - N_- +
  \tilde{N}_-/\kappa^3 \right) \\ - \frac{\kappa(\beta_1 -
  (1+\kappa)\alpha_2)}2\left( \left(N_- - \tilde{N}_-/\kappa \right)^2
  - N_+ - \tilde{N}_+/\kappa^2 \right),
\end{multline*}
\end{cor}
where
\begin{equation*}
  N_\pm = N\pm M,\quad \tilde{N}_\pm = \tilde{N}\pm \tilde{M}.
\end{equation*}

Using this identity we can straightforwardly generalise our solution
method and obtain many different formulae for eigenfunctions of the
deformed Calogero-Sutherland operators \eqref{D}: for fixed particle
numbers $N$ and $\tilde{N}$ one is free to choose $M$ and $\tilde{M}$
arbitrarily and each choice gives a family of eigenfunctions labelled
by integer vectors $\mathbf{n}\in \mathbb{Z}^{M+\tilde{M}}$. As
discussed in Section~\ref{sec42}, we obtain in this way a number of
series representations for each eigenfunction, a fact which is
interesting already in the standard (non-deformed) case $\tilde{N} =
0$.

\subsection{Case II}
In order to put the general discussion in the previous two sections
into perspective we consider in some detail the special
case II; see Tables \ref{table1} and \ref{table2}. The Schr\"odinger
operator \eqref{HN} is in this case given by
\begin{equation*}
  H_N = -\sum_{j=1}^N\partial_{x_j}^2 + 2\kappa(\kappa-1)\sum_{j<k}\frac{1}{4\sin^2\left(\frac{x_j-x_k}{2}\right)}.
\end{equation*}
As we mentioned in the introduction, this Schr\"odinger operator was
introduced and studied by Sutherland \cite{Su1,Su2}, and its
eigenfunctions are given by the so-called Jack polynomials. In this
special case the identity in Corollary \ref{cor2} is equivalent to a
well-known and important identity in the theory of Jack polynomials,
and Corollary \ref{cor4} gives a natural generalisation of this
identity to the deformed case.  More precisely, we have the following:

\begin{cor}
Let
\begin{equation*}
\begin{split}
  F(\mathbf{x},\mathbf{y}) =
  \prod_{j=1}^Nw_j^\kappa\frac{\prod_{j<k}(z_k-z_j)^{\kappa}
    (w_k-w_j)^{\kappa} }{\prod_{j,k}(w_k-z_j)^{\kappa}}
\end{split}
\end{equation*} 
with $z_j=e^{ix_j}$ and $w_j=e^{iy_j}$. Then
\begin{equation}\label{IdCaseII} 
  \left( H_{N}(\mathbf{x}) - H_{N}(\mathbf{y})
 \right) F(\mathbf{x},\mathbf{y}) = 0,
\end{equation}
where the arguments indicate that the Schr\"odinger operator acts in the 
variables $\mathbf{x}$ and $\mathbf{y}$, respectively.
\end{cor}

As mentioned above, the identity \eqref{IdCaseII}, and in particular
the function $F$, is directly related with the theory of Jack
polynomials.  In order to make this precise we observe that
\begin{equation*}
  F(\mathbf{z},\mathbf{w}) = \prod_{j<k}(z_k - z_j)^\kappa (1/w_j - 1/w_k)^\kappa \Pi(\mathbf{z},\mathbf{w})
\end{equation*}
with the function $\Pi$ given by
\begin{equation*}
  \Pi(\mathbf{z},\mathbf{w}) = \frac{1}{\prod_{j,k}(1-z_j/w_k)^\kappa}.
\end{equation*}
It is well known that this latter function $\Pi$ has the following
expansion in the monic Jack polynomials $\mathcal{J}_{\boldsymbol{\lambda}}$:
\begin{equation}\label{PiExpInJacks}
  \Pi(\mathbf{z},\mathbf{w}) = \sum_{\boldsymbol{\lambda}} b_{\boldsymbol{\lambda}} \mathcal{J}_{\boldsymbol{\lambda}}(\mathbf{z})\mathcal{J}_{\boldsymbol{\lambda}}(\mathbf{w}^{-1})
\end{equation}
for some (explicitly known) coefficients $b_{\boldsymbol{\lambda}}$,
and where $\mathbf{w}^{-1} = (1/w_1,\ldots,1/w_N)$; see, e.g., Sections
VI.4 and VI.10 in \cite{MacD}. We recall that the monic Jack
polynomials are usually denoted $P_{\boldsymbol{\lambda}}$ rather than
$\mathcal{J}_{\boldsymbol{\lambda}}$. We have used the latter notation
in order to avoid clashes with notation used in later sections. Up to
degeneracies in the spectrum of the Schr\"odinger operator $H_N$, this
expansion can be deduced as a consequence of the identity
\eqref{IdCaseII}. To establish this fact we let $\Psi_0(\mathbf{z}) =
\prod_{j<k}(z_k-z_j)^\kappa$, and similarly for
$\Psi_0(\mathbf{w}^{-1})$. It is well known that the functions
$\Psi_0(\mathbf{z})\mathcal{J}_{\boldsymbol{\lambda}}(\mathbf{z})$ are
eigenfunctions of $H_N$; this can for example be inferred from results
of Stanley \cite{Stanley}. Since $H_N(\mathbf{y})$ is invariant under
the substitution of $-y_j$ for $y_j$ which maps $w_j$ to $1/w_j$, the
functions
$\Psi_0(\mathbf{w}^{-1})\mathcal{J}_{\boldsymbol{\lambda}}(\mathbf{w}^{-1})$
are eigenfunctions of the same Schr\"odinger operator $H_N$, and with
the same eigenvalues. We observe that $\Pi(\mathbf{z},\mathbf{w})$ is
invariant under permutations of both the variables $\mathbf{z}$ as
well as the variables $\mathbf{w}$. It follows that it has an
expansion in Jack polynomials of the form
\begin{equation*}
  \Pi(\mathbf{z},\mathbf{w}) = \sum_{\boldsymbol{\lambda}} Q_{\boldsymbol{\lambda}}(\mathbf{z})\mathcal{J}_{\boldsymbol{\lambda}}(\mathbf{w}^{-1}),
\end{equation*}
where the functions $Q_{\boldsymbol{\lambda}}$ are to be determined. Since
$\Psi_0(\mathbf{w}^{-1})\mathcal{J}_{\boldsymbol{\lambda}}(\mathbf{w}^{-1})$ is an
eigenfunction of $H_N$, the identity \eqref{IdCaseII} implies that
also $\Psi_0(\mathbf{z})Q_{\boldsymbol{\lambda}}(\mathbf{z})$ is an eigenfunction of
$H_N$, and with the same eigenvalue. It follows that, unless this 
eigenvalue is degenerate, $Q_{\boldsymbol{\lambda}}(\mathbf{z})$ is proportional to $\mathcal{J}_{\boldsymbol{\lambda}}(\mathbf{z})$. We mention that although such degeneracies do 
occur they are quite rare in this case.

The identities obtained in the previous section thus suggest that
there exist expansions, similar to the one given above, for each
function $F$ corresponding to a Schr\"odinger operator \eqref{HN} or
a deformed counterpart \eqref{D}. However, in most cases the
Schr\"odinger operator $H_N$, as well as its deformed counterpart $H_{N,\tilde{N}}$, is not invariant under a change of
coordinates mapping $w_j$ to $1/w_j$. This explains the appearance of
the operator $H_N^{(-)}$ and the constant $C_N$ in Corollary
\ref{cor2}, as well as that of $H^{(-)}_{M,\tilde{M}}$ and $C_{N,\tilde{N},M,\tilde{M}}$ in Corollary \ref{cor4}.

\section{Eigenfunctions of Calogero-Sutherland type models}\label{sec3} 
In this section we construct our series representations of the
eigenfunctions of the Schr\"odinger operators \eqref{HN}. These series
representations are in terms of the functions $f_{\mathbf{n}}$ defined
in \eqref{fn0}. As we will see, the identities deduced in the previous
section play a central role in this construction. These series
representation generalises those obtained in \cite{Lang} for the
special case~II and in \cite{HL} for I and IV (c.f.\ Table
\ref{table1}).

\subsection{Construction of eigenfunctions}\label{sec31}
We fix the polynomials $\alpha$, $\beta$ in \eqref{ab} and the
coupling constant $\kappa>0$ and consider the resulting Schr\"odinger
operator $H_N$ as defined by \eqref{HN}, \eqref{V1}, \eqref{R2} and
\eqref{W}. We observe that the eigenfunctions in \eqref{Psin} are
completely determined by the functions $P_{\mathbf{n}}$ in $N$
variables $\mathbf{z}$, and we will refer to the latter as
\textit{reduced eigenfunctions} of $H_N$. We also observe that
$P_{\mathbf{n}}$ is a reduced eigenfunction of $H_N$ if and only if it
is an eigenfunction of the differential operator
\begin{equation}\label{redHN}
  \tilde{H}_N := \Psi^{-1}_0(H_N - E_0)\Psi_0 =
  -\sum_{j=1}^N\left(\alpha(z_j)\partial_{z_j}^2 +
  \beta(z_j)\partial_{z_j}\right) - 2\kappa\sum_{j\neq
  k}\frac{\alpha(z_j)}{z_j - z_k}\partial_{z_j},
\end{equation}
where the last equality follows by a straightforward computation using
the fact that $\Psi_0$ is an eigenstate of $H_N$ with eigenvalue $E_0$
and the identities in \eqref{th}, \eqref{W}, and \eqref{R2}. Using the
identity in Corollary \ref{cor2} it is now straightforward to
compute the action of this differential operator on the polynomials
$f_{\mathbf{n}}$. In order to state this action in a simple form, and
also to facilitate a discussion of its implications, we first
introduce some convenient notation. An important ingredient in our
construction is the following partial ordering of integer vectors
$\mathbf{n},\mathbf{m}\in\mathbb{Z}^N$:
\begin{equation}\label{ordering} 
  \mathbf{m}\preceq\mathbf{n} \Leftrightarrow m_j + \cdots + m_N\leq
  n_j + \cdots + n_N, \quad \forall j=1,2,\ldots, N.
\end{equation}
That this is only a partial ordering is easily seen, e.g., $(522)$ and
$(441)$ are incomparable. The algebra of symmetric polynomials
$P(\mathbf{z})$ in $N$ variables $\mathbf{z} = (z_1,\ldots,z_N)$ with
complex coefficients is denoted $\Lambda_N$. The linear subspace
consisting of the homogeneous symmetric polynomials of degree $n$,
together with the zero polynomial, is denoted by $\Lambda^n_N$. We
also denote as $\mathbf{e}_j$ the standard basis in $\mathbb{Z}^N$,
i.e., $(\mathbf{e}_j)_k=\delta_{jk}$ for all $j,k=1,2,\ldots,N$, and
let
\begin{equation*}
  \mathbf{E}_{j,k}^{p,\nu} = (1 - \nu)\mathbf{e}_j + (1 - p + \nu)\mathbf{e}_k.
\end{equation*}
As before,
\begin{equation}
  \label{absvn}
  |\mathbf{n}| := n_1+ \ldots + n_N.
\end{equation}
For simplicity of notation we shall to each $\mathbf{n}\in
\mathbb{Z}^N$ associate the shifted integer vector $\mathbf{n}^+ =
(n^+_1,\ldots,n^+_N)$ with
\begin{equation}\label{npj}
  n^+_j = n^{\phantom\dag}_j + \kappa(N+1-j).
\end{equation}
We are now ready to state and prove the following:

\begin{lemma}\label{actionLemma}
For each $\mathbf{n}\in\mathbb{Z}^N$,
\begin{multline}\label{action}
  \tilde{H}_N f_{\mathbf{n}} = (E_{\mathbf{n}} - E_0)f_{\mathbf{n}}\\
  - \sum_{j=1}^N\big(\alpha_1 n^+_j(n^+_j-1) + (\beta_0- (1+\kappa)
  \alpha_1)(n^+_j - 1)\big)f_{\mathbf{n}- \mathbf{e}_j}\\ -
  \alpha_0\sum_{j=1}^N (n^+_j-1)(n^+_j-2)f_{\mathbf{n}- 2
  \mathbf{e}_j}\\ + \kappa(\kappa-1) \sum_{j<k}\sum_{p=0}^2
  \sum_{\nu=1}^\infty \alpha_p (2\nu-p)f_{\mathbf{n} -
  \mathbf{E}_{j,k}^{p,\nu}}.
\end{multline}
\end{lemma}

\begin{proof} 
Note that the function $F$ in Corollary~\ref{cor2} can be written as
follows:
\begin{equation*}
  F(\mathbf{x},\mathbf{y}) = \Psi_0(\mathbf{x}) G(\mathbf{y})
  \prod_{j=1}^N w_j^{-\kappa(N+1-j)}
  \frac{\prod_{j<k}(1-w_j/w_k)^{\kappa}}{\prod_{j,k}
  (1-z_j/w_k)^\kappa}
\end{equation*}
with
\begin{equation*}
  G(\mathbf{y}) := \prod_{j=1}^N \psi_{0,-1}(y_j).
\end{equation*}
The definition~\eqref{fn0} of the polynomials $f_{\mathbf{n}}$ thus
implies
\begin{equation*}
  F(\mathbf{x},\mathbf{y}) = \Psi_0(\mathbf{x})G(\mathbf{y})
  \sum_{\mathbf{n}\in\mathbb{Z}^N} f_{\mathbf{n}}(\mathbf{z})
  \mathbf{w}^{-\mathbf{n}^+} . 
\end{equation*}
It follows from Corollary~\ref{cor2} that
\begin{equation}\label{key} 
\sum_{\mathbf{n}\in\mathbb{Z}^N}(\tilde{H}_N
  f_{\mathbf{n}}(\mathbf{z}))\mathbf{w}^{-\mathbf{n}^+} =
  \sum_{\mathbf{n}\in\mathbb{Z}^N}f_{\mathbf{n}}(\mathbf{z})(
  \bar{H}_N^{(-)} + C_N) \mathbf{w}^{-\mathbf{n}^+}
\end{equation}
where
\begin{equation*}
  \bar{H}_N^{(-)} := G^{-1}(\mathbf{y})H_N^{(-)}G(\mathbf{y}).
\end{equation*}
Using~\eqref{th}, \eqref{W}, and \eqref{R2}, as well as the definition
of $H^{(-)}_N$, we find that
\begin{equation*}
  \bar{H}_N^{(-)} = -\sum_{j=1}^N \left(\alpha(w_j)\partial_{w_j}^2 +
  \beta_{-1}(w_j) \partial_{w_j}\right) + \kappa(\kappa-1) \sum_{j<k}
  \left(\frac{\alpha(w_j)+\alpha(w_k)}{(w_j-w_k)^2} - \alpha_2 \right)
\end{equation*}
with
\begin{equation*}
  \beta_{-1}(w) = -\beta(w) + (1+\kappa)\alpha'(w).
\end{equation*}
Recalling~\eqref{ab} we expand the interaction terms in Laurent series
and obtain
\begin{equation*}
\begin{split}
  \frac{\alpha(w_j)+\alpha(w_k)}{(w_j-w_k)^2} -\alpha_2 &= \frac{2
  \alpha_2 w_jw_k + \alpha_1 (w_j+w_k) + 2
  \alpha_0}{w_k^2(1-w_j/w_k)^2}\\ &= \sum_{p=0}^2 \sum_{\nu=1}^\infty
  \alpha_p (2\nu-p) \frac{w_j^{\nu-1}}{w_k^{\nu+1-p}}
\end{split}
\end{equation*}
valid for $|w_N|>|w_{N-1}|>\cdots > |w_1|>\max_k(|z_k|)$. It is now
straightforward to compute the r.h.s.\ of \eqref{key}, and by
comparing the coefficients of $\mathbf{w}^{-\mathbf{n}^+}$ on both
sides of the resulting equation we obtain \eqref{action} with
\begin{equation*}
  E_{\mathbf{n}} = -\sum_{j=1}^N [ \alpha_2 n^+_j (n^+_j+1) + (\beta_1
  - 2(1+\kappa)\alpha_2)n^+_j ] + C_N.
\end{equation*}
By straightforward computations we find that the latter coincides with
$E_{\mathbf{n}}$ in \eqref{En} and \eqref{E0}.
\end{proof} 

It is clear that the action of $\tilde{H}_N$ on the polynomials
$f_{\mathbf{n}}$ is triangular in the sense that $\tilde{H}_N
f_{\mathbf{n}}$ is a linear combination of polynomials
$f_{\mathbf{m}}$ with $\mathbf{m}\preceq\mathbf{n}$. In fact,
depending on the specific choice of $\alpha$ and $\beta$, the set of
possible $\mathbf{m}$ is in many cases significantly smaller. In any
case, this triangular structure suggests that there exist
eigenfunctions, with corresponding eigenvalues $E_{\mathbf{n}} - E_0$,
of the differential operator $\tilde{H}_N$ which are of the form
\begin{equation}\label{eigFuncAnsatz}
  P_{\mathbf{n}} = f_{\mathbf{n}} + \sum_{\mathbf{m}}
  u_{\mathbf{n}}(\mathbf{m})f_{\mathbf{m}}
\end{equation}
where the sum is over integer vectors
$\mathbf{m}\prec\mathbf{n}$. Indeed, if we take this as an ansatz then
we find that this is the case if and only if the coefficients satisfy
the recursion relation
\begin{multline}\label{recursion}
  (E_{\mathbf{n}} - E_{\mathbf{m}})u_{\mathbf{n}}(\mathbf{m}) =
  -\sum_{j=1}^N m_j^+(\beta_0 + \alpha_1(m_j^+ -
  \kappa))u_{\mathbf{n}}(\mathbf{m}+\mathbf{e}_j)\\ -
  \alpha_0\sum_{j=1}^N m_j^+(m_j^+ +
  1)u_{\mathbf{n}}(\mathbf{m}+2\mathbf{e}_j) + \kappa(\kappa -
  1)\sum_{j<k}\sum_{p=0}^2 \alpha_p
  (2\nu-p)u_{\mathbf{n}}(\mathbf{m}+\mathbf{E}^{p,\nu}_{j,k}).
\end{multline}
Since $u_{\mathbf{n}}(\mathbf{n})$ is fixed to one, it is clear that
if $E_{\mathbf{n}} - E_{\mathbf{m}}\neq 0$ for all $\mathbf{m}$ which
appear in \eqref{eigFuncAnsatz}, then this recursion relation uniquely
determines the remaining coefficients
$u_{\mathbf{n}}(\mathbf{m})$. This condition of non-degeneracy on the
eigenvalues is generically satisfied, but there exist special cases
and specific integer vectors $\mathbf{n}$ for which it fails to hold
true; see Remark \ref{rmk32} for a further discussion of this point.

\subsection{Case II} 
In order to draw attention to some of the specific features of the
action of $\tilde{H}_N$ on the polynomials $f_{\mathbf{n}}$ we
consider in some detail the special case II; see Tables \ref{table1}
and \ref{table2}. On the one hand this is in many ways the simplest
non trivial case, and our construction of reduced eigenfunctions is
therefore particularly simple in this case, but, on the other hand, it
already contains many of the key features of the general case. For the
special case II we have $\alpha(z) = -z^2$ and $\beta(z) = -z$, and
consequently that
\begin{equation*}
  \tilde{H}_N f_{\mathbf{n}} = (E_{\mathbf{n}} - E_0)f_{\mathbf{n}} - 2\kappa(\kappa-1)\sum_{j<k}\sum_{\nu=1}^\infty \nu f_{\mathbf{n}-\nu(\mathbf{e}_k-\mathbf{e}_j)}.
\end{equation*}
Clearly, the right hand side now only contains polynomials
$f_{\mathbf{m}}$ such that $|\mathbf{m}| = |\mathbf{n}|$. It is
interesting to compare this expression with the corresponding action
of $\tilde{H}_N$ on the monomial symmetric polynomials
\begin{equation}\label{monomials}
  m_{\boldsymbol{\lambda}}(\mathbf{z}) := \sum_P z_1^{\lambda_{P(1)}}\cdots
  z_N^{\lambda_{P(N)}}
\end{equation}
with $\boldsymbol{\lambda}$ a partition of length
$\ell(\boldsymbol{\lambda})\leq N$, and where the sum extends over all
distinct permutations $P$ of the parts $\lambda_j$ of
$\boldsymbol{\lambda}$. To this end we let $\lfloor x\rfloor$ denote
the integer part of $x\in\mathbb{R}$. In addition, for any
$\mathbf{n}\in\mathbb{N}_0^N$ we identify $m_{\mathbf{n}}$ with the
unique $m_{\boldsymbol{\lambda}}$ such that $\mathbf{n} =
P(\boldsymbol{\lambda})$ for some permutation $P$. Using the identity
\begin{equation*}
  \frac{1}{x-y}(x^2\partial_x - y^2\partial_y)(x^ny^m + x^my^n) = (n - m)\sum_{k=0}^{n-m}x^{n-k}y^{m+k} + 2m x^ny^m,
\end{equation*}
valid for any non-negative integers $n$ and $m$ such that $n\geq m$,
it is now easy to verify that
\begin{equation*}
  \tilde{H}_N m_{\boldsymbol{\lambda}} = (E_{\boldsymbol{\lambda}} - E_0)m_{\boldsymbol{\lambda}} + 2\kappa \sum_{j<k}(\lambda_j-\lambda_k)\sum_{\nu=0}^{\lfloor (\lambda_j-\lambda_k)/2\rfloor} m_{\boldsymbol{\lambda}+\nu(\mathbf{e}_k-\mathbf{e}_j)}.
\end{equation*}
We note, in particular, that the action of $\tilde{H}_N$ on the
polynomials $f_{\mathbf{n}}$ does not contain an explicit dependence
on the integer vector $\mathbf{n}$. In contrast, the action on the
symmetric monomials $m_{\boldsymbol{\lambda}}$ do contain an explicit
dependence on the partition $\boldsymbol{\lambda}$. It is also
interesting to note that the corresponding recursion relation
\eqref{recursion} can in fact be solved to yield the following
explicit series representation for the eigenfunctions of
$\tilde{H}_N$:
\begin{equation*}
\begin{split}
  P_\mathbf{n} &= f_{\mathbf{n}} + \sum_{s=1}^\infty
  (2\kappa(1-\kappa))^s \sum_{j_1<k_1}\sum_{\nu_1=1}^\infty
  \nu_1\times\cdots\times\sum_{j_s<k_s}\sum_{\nu_s=1}^\infty \nu_s\\
  &\quad \times\prod_{r=1}^s \left(E_{\boldsymbol{\lambda}} -
    E_{\boldsymbol{\lambda}-\sum_{t=r}^s\nu_t(\mathbf{e}_{k_t}-\mathbf{e}_{j_t})}\right)^{-1}
  f_{\boldsymbol{\lambda}-\sum_{r=1}^s\nu_r(\mathbf{e}_{k_r}-\mathbf{e}_{j_r})};
\end{split}
\end{equation*}
see \cite{Lang3} for further details. It is important to note that the
fact that $f_{\mathbf{n}}$ is non-zero only if $\mathbf{n}\succeq 0$
(see Corollary \ref{cor31}) implies that this series only contains a
finite number of terms, and thus is well-defined. Even though it can
be done in principle, it seems harder to solve the recursion relations
resulting from the use of the monomial symmetric polynomials. Detailed
discussions of the above construction of series representations for
the reduced eigenfunctions of the Schr\"odinger operator \eqref{HN} in
the cases I and IV can be found in \cite{HL,Hal1}. We also mention
that it is possible to write down explicit formulae for the reduced
eigenfunctions of the Schr\"odinger operator \eqref{HN} in the general
case which, however, are somewhat involved.\footnote{The interested
  reader can find these formulae in the first arXiv version of the
  present paper; see http://arxiv.org/abs/math-ph/0703090v1.}

\subsection{Completeness of the reduced eigenfunctions}\label{sec33}
In Section \ref{sec31} we constructed, under a certain condition of
non-degeneracy on the corresponding eigenvalues, for each integer
vector $\mathbf{n}\in\mathbb{Z}^N$ a reduced eigenfuntion
$P_{\mathbf{n}}$ of the Schr\"odinger operator $H_N$. In this section
we show that if we restrict attention to integer vectors $\mathbf{n} =
\boldsymbol{\lambda}$ for some partition $\boldsymbol{\lambda}$, then
the corresponding reduced eigenfunctions form a linear basis for
$\Lambda_N$. The idea of the proof is to first establish that the
expansion of the symmetric polynomials $f_{\mathbf{n}}$ in terms of a
particular set of symmetric polynomials $g_{\boldsymbol{\lambda}}$,
known to constitute such a basis, has a particular triangular
structure. This implies that the $f_{\boldsymbol{\lambda}}$,
labelled by the partitions $\boldsymbol{\lambda}$ of length
$\ell(\boldsymbol{\lambda})\leq N$, form a linear basis for
$\Lambda_N$. We then obtain a similar result for the reduced
eigenfunctions $P_{\boldsymbol{\lambda}}$ by showing that they are, in
the same sense, triangular in the $f_{\boldsymbol{\lambda}}$.

We proceed to recall that the polynomials $g_{\boldsymbol{\lambda}}$
in question, sometimes referred to as the modified complete symmetric
polynomials, can be defined by the expansion
($\mathbf{w}^{-1}=(w_1^{-1},\ldots,w_N^{-1})$)
\begin{equation}\label{PiExpInGs}
  \Pi(\mathbf{z},\mathbf{w}) = \sum_{\boldsymbol{\lambda}}
  g_{\boldsymbol{\lambda}}(\mathbf{z}) m_{\boldsymbol{\lambda}}(
  \mathbf{w}^{-1})
\end{equation}
valid for $\min_k(|w_k|)>\max_j(|z_j|)$, and where the sum extends
over all partitions $\boldsymbol{\lambda}$ of length
$\ell(\boldsymbol{\lambda})\leq N$. It is well known that the
$g_{\boldsymbol{\lambda}}$ constitute a linear basis for $\Lambda_N$;
see, e.g., Section~VI.10 in \cite{MacD}. Using this fact we now
establish the following relation between the $f_{\mathbf{n}}$ and the
$g_{\boldsymbol{\lambda}}$, which, in particular, implies that the
$f_{\mathbf{n}}$ indeed are well-defined symmetric polynomials:

\begin{lemma}\label{ftriangLemma}
Let $\mathbf{n}\in\mathbb{Z}^N$. Then
\begin{equation*}
  f_{\mathbf{n}} = \sum_{\boldsymbol{\mu}}
  a_{\mathbf{n}\boldsymbol{\mu}} g_{\boldsymbol{\mu}}
\end{equation*}
for some coefficients $a_{\mathbf{n}\boldsymbol{\mu}}$ and where the
sum is over partitions $\boldsymbol{\mu}\preceq\mathbf{n}$ such that
$|\boldsymbol{\mu}| = |\mathbf{n}|$. Moreover, if $\mathbf{n} =
\boldsymbol{\lambda}$ for some partition $\boldsymbol{\lambda}$ of
length $\ell(\boldsymbol{\lambda})\leq N$, then
$a_{\boldsymbol{\lambda}\boldsymbol{\lambda}} = 1$.
\end{lemma}

\begin{proof}
It follows from the definitions of the $g_{\boldsymbol{\lambda}}$ and
$f_{\mathbf{n}}$ that
\begin{equation*}
\begin{split}
 \sum_{\mathbf{n}\in\mathbb{Z}^N}f_{\mathbf{n}}(\mathbf{z})
 \mathbf{w}^{-\mathbf{n}} &=
 \prod_{j<k}(1-w_j/w_k)^\kappa\sum_{\mathbf{m}\in\mathbb{N}_0^N}
 g_{p(\mathbf{m})}(\mathbf{z})\mathbf{w}^{-\mathbf{m}}\\& =
 \sum_{\mathbf{m}\in\mathbb{N}_0^N}g_{p(\mathbf{m})}(\mathbf{z})
 \prod_{j<k}\sum_{q_{jk}=0}^{\infty}(-1)^{q_{jk}}
 \binom{\kappa}{q_{jk}}\mathbf{w}^{-\mathbf{m} +
 \sum_{j<k}q_{jk}(\mathbf{e}_j- \mathbf{e}_k)}
\end{split}
\end{equation*}
where $p(\mathbf{m})$ denotes the unique partition obtained by
permuting the components of the integer vector $\mathbf{m}$, e.g.,
$p(2,4,5,2)=(5,4,2,2)$.  Comparing the coefficients of
$\mathbf{w}^{-\mathbf{n}}$ on both sides of this identity we find that
\begin{equation}\label{jacobiTrudi}
  f_{\mathbf{n}} =
  \prod_{j<k}\sum_{q_{jk}=0}^{\infty}(-1)^{q_{jk}}\binom{\kappa}{q_{jk}}
  g_{p(\mathbf{n}+\sum_{j<k}q_{jk}(\mathbf{e}_j-\mathbf{e}_k))}.
\end{equation}
We thus obtained a representation of $f_{\mathbf{n}}$ as a linear
superposition of polynomials $g_{\boldsymbol{\mu}}$ where
$\boldsymbol{\mu}=p(\mathbf{m})$ for
\begin{equation*}
  \mathbf{m} = \mathbf{n} + \sum_{j<k}q_{jk}(\mathbf{e}_j -
  \mathbf{e}_k)
\end{equation*}
with non-negative integers $q_{jk}$. Obviously,
$|\boldsymbol{\mu}|=|\mathbf{m}| = |\mathbf{n}|$ for such
$\boldsymbol{\mu}$.  To complete the proof we thus only need to show
that $\boldsymbol{\mu} \preceq\mathbf{n}$.  For that we observe that
$\mathbf{m}\preceq\mathbf{n}$. Moreover, since $\boldsymbol{\mu}$ is a
partition, $\mu_1\geq\mu_2\geq\cdots\geq\mu_N$, and the definition of
the ordering \eqref{ordering} thus implies that ${\boldsymbol{\mu}}
\preceq P{\boldsymbol{\mu}}$ for any permutation $P$ of the parts of
${\boldsymbol{\mu}}$ and, in particular, ${\boldsymbol{\mu}} \preceq
\mathbf{m}$.
\end{proof}

\begin{remark}
  It is interesting to note that in the particular case $\kappa=1$ the
  result in \eqref{jacobiTrudi} is equivalent to the so-called
  Jacobi-Trudi identity (originally due to Jacobi \cite{Jacobi}),
  which implies that the well-known Schur polynomials
  $s_{\boldsymbol{\lambda}}$ are identical with the polynomials
  $f_{\boldsymbol{\lambda}}$ for $\kappa=1$. The proof of this fact
  can be obtained as a simple consequence of well-known arguments
  which can be found in Macdonald's book \cite{MacD}; see the proof of
  (3.4'') and the preceding discussion in Chapter~I.
\end{remark}

The fact that each $g_{\boldsymbol{\mu}}$ is a symmetric homogeneous
polynomial of degree $|{\boldsymbol{\mu}}|$ now implies the following:

\begin{cor}
\label{cor31}
The function $f_{\mathbf{n}}$, $\mathbf{n}\in\mathbb{Z}^N$, is
non-zero only if $\mathbf{n}\succeq \mathbf{0}$. In that case it is
a symmetric homogeneous polynomial of degree $|\mathbf{n}|$.
\end{cor}

Let $n$ be a non-negative integer and consider the two sets of
polynomials $g_{\boldsymbol{\lambda}}$ and $f_{\boldsymbol{\lambda}}$
which are both indexed by those partitions $\boldsymbol{\lambda}$ of
$n$ which are of length $\ell(\boldsymbol{\lambda})\leq N$. Let $K =
(K_{\boldsymbol{\lambda} \boldsymbol{\mu}} )$ denote the transition
matrix from the $f_{\boldsymbol{\lambda}}$ to the
$g_{\boldsymbol{\lambda}}$, defined by the equalities
$f_{\boldsymbol{\lambda}} = \sum_{\boldsymbol{\mu}}
K_{\boldsymbol{\lambda}\boldsymbol{\mu}}g_{\boldsymbol{\mu}}$. Given a
partition $\boldsymbol{\lambda}$, Lemma~\ref{ftriangLemma} implies
that $K_{\boldsymbol{\lambda}{\boldsymbol{\mu}}}=0$ unless
${\boldsymbol{\mu}}\preceq\boldsymbol{\lambda}$ and that
$K_{\boldsymbol{\lambda}\boldsymbol{\lambda}} = 1$. It is readily
verified that the inverse of such a matrix always exist and is of the
same form; see, e.g., Section~I.6 in \cite{MacD}. It follows that
\begin{equation}\label{gtriang}
  g_{\boldsymbol{\lambda}} = f_{\boldsymbol{\lambda}} +
  \sum_{\boldsymbol{\mu}}(K^{-1})_{\boldsymbol{\lambda}{\boldsymbol{\mu}}}f_{\boldsymbol{\mu}}
\end{equation}
where $(K^{-1})_{\boldsymbol{\lambda}{\boldsymbol{\mu}}}$ are the
elements in the matrix inverse to $K$ and the sum is over partitions
$\boldsymbol{\mu} \prec \boldsymbol{\lambda}$ of $n$. We thus obtain
the following:

\begin{prop}\label{basisProp}
For any non-negative integer $n$, the polynomials
$f_{\boldsymbol{\lambda}}$, labelled by those partitions
$\boldsymbol{\lambda}$ of $n$ which are of length
$\ell(\boldsymbol{\lambda})\leq N$, form a linear basis for
$\Lambda^n_N$.
\end{prop}

Suppose now that the reduced eigenfunctions $P_{\mathbf{n}}$ of the
form \eqref{eigFuncAnsatz} exist for all integer vectors
$\mathbf{n}\in\mathbb{Z}^N$ such that $\mathbf{n} =
\boldsymbol{\lambda}$ for some partition $\boldsymbol{\lambda}$. From
Lemma~\ref{ftriangLemma} and \eqref{gtriang} we have that any such
reduced eigenfunction has a series expansion of the form
\begin{equation*}
  P_{\boldsymbol{\lambda}} = f_{\boldsymbol{\lambda}} + \sum_{\boldsymbol{\mu}}
  v_{\boldsymbol{\lambda}{\boldsymbol{\mu}}}f_{\boldsymbol{\mu}}
\end{equation*}
for some coefficients $v_{\boldsymbol{\lambda}{\boldsymbol{\mu}}}$,
and where the sum is over partitions
${\boldsymbol{\mu}}\prec\boldsymbol{\lambda}$. Fix a partition
$\boldsymbol{\lambda}$ of length $\ell(\boldsymbol{\lambda})\leq N$
and consider the transition matrix from the reduced eigenfunctions
$P_{\boldsymbol{\mu}}$ to the polynomials $f_{\boldsymbol{\mu}}$, both
indexed by the partitions ${\boldsymbol{\mu}}$ of length
$\ell(\boldsymbol{\mu})\leq N$ such that
${\boldsymbol{\mu}}\preceq\boldsymbol{\lambda}$. Observe that its
entries are non-zero only if
${\boldsymbol{\mu}}\preceq\boldsymbol{\lambda}$ and that all diagonal
entries are equal to one. It is clear from the arguments leading up to
Proposition~\ref{basisProp} that this transition matrix has a
well-defined inverse of the same form. It follows that each polynomial
$f_{\boldsymbol{\lambda}}$ is a linear combination of reduced
eigenfunctions $P_{\boldsymbol{\mu}}$ with ${\boldsymbol{\mu}}\preceq
\boldsymbol{\lambda}$. Proposition~\ref{basisProp} thus implies the
following:

\begin{prop}\label{prop32}
If the reduced eigenfunctions $P_{\boldsymbol{\lambda}}$ in
\eqref{eigFuncAnsatz} exist for all partitions
$\boldsymbol{\lambda}$ of length $\ell(\boldsymbol{\lambda})\leq N$
then they form a linear basis for $\Lambda_N$.
\end{prop}

\begin{remark}\label{rmk32}
As discussed after Equation \eqref{recursion}, a reduced
eigenfunction of the form \eqref{eigFuncAnsatz} exist if
$E_{\mathbf{n}} - E_{\mathbf{m}}\neq 0$ for all $\mathbf{m}$ which
appear in \eqref{eigFuncAnsatz}. It is easy to see that this
condition of non-degeneracy holds true if $\alpha_2 = 0$ (and
$\beta_1\neq 0$), including the special cases I and IV (see, e.g.,
\cite{HL}). It is also known to hold for all integer vectors
$\mathbf{n}\in\mathbb{Z}^N$ such that $\mathbf{n} =
\boldsymbol{\lambda}$ for some partition $\boldsymbol{\lambda}$ in
the special case II (see, e.g., \cite{Lang3}), assuming that
$\kappa>0$. It is interesting to note that if $\alpha_2\neq 0$ and
if one allow $\kappa$ to be negative then there exist parameter
values and integer vectors $\mathbf{n}$ such that $E_{\mathbf{n}} =
E_{\mathbf{m}}$ for some integer vector $\mathbf{m}$ which appear in
\eqref{eigFuncAnsatz}. We stress, however, that for generic
parameter values and integer vectors $\mathbf{n}$ this condition of
non-degeneracy on the eigenvalues is satisfied.
\end{remark}

\section{Eigenfunctions of deformed Calogero-Sutherland type models}
\label{sec4}
In this section we sketch the generalisation of our construction of
eigenfunctions from the previous section to the deformed
Calogero-Sutherland operators $H_{N,\tilde{N}}$ \eqref{D}, emphasising
the features which are new. Just as the construction in the previous
section was based on Corollary \ref{cor2}, the discussion below is
based on Corollary \ref{cor4}. We recall that the identity in this
latter corollary relates the operator $H_{N,\tilde{N}}$ to an operator
of the same type, but with particle numbers $M$ and $\tilde{M}$. We
stress that these particle numbers can be chosen freely, and that, as
we will discuss in Section \ref{sec42}, the complexity of the
resulting series representation of a given eigenfunction can be highly
dependent on the specific choice. As described already in the
beginning of Section~\ref{sec1}, this can be used to construct simpler
explicit formulae for many of the eigenfunctions already in the
special case $\tilde{N}=0$.

We proceed to describe the structure of the eigenfunctions of the
deformed Calogero-Sutherland operators \eqref{D}. Following Sergeev
and Veselov \cite{SV2} we let $\Lambda_{N,\tilde{N},\kappa}$ be the
algebra of polynomials $P(\mathbf{z},\mathbf{\tilde{z}})$ in two sets
of independent variables $\mathbf{z} = (z_1,\ldots,z_N)$ and
$\mathbf{\tilde{z}} = (\tilde{z}_1,\ldots,\tilde{z}_{\tilde{N}})$ with
complex coefficients such that they are separately symmetric in the
variables $\mathbf{z}$ and $\mathbf{\tilde{z}}$ and, furthermore, obey
the condition
\begin{equation*}
(\partial_{z_j} + \kappa\partial_{\tilde{z}_J}) P(\mathbf{z},
  {\mathbf{\tilde{z}}}) \arrowvert_{z_j=\tilde{z}_J} = 0
\end{equation*}
for all $j = 1,\ldots,N$ and $J = 1,\ldots,\tilde{N}$. We will show
that the deformed Calogero-Sutherland operators \eqref{D} have
eigenfunctions which are naturally labelled by partitions
$\boldsymbol{\lambda} = (\lambda_1,\lambda_2,\ldots)$ in the so-called
fat $(N,\tilde{N})$-hook, i.e., $\lambda_{N+1}\leq\tilde{N}$, and are
of the form
\begin{equation}
  \label{PsinD}
  \Psi_{\boldsymbol{\lambda}}(\mathbf{x}, \mathbf{\tilde{x}}) =
  \Psi_0(\mathbf{x},{\mathbf{\tilde{x}}}) P_{\boldsymbol{\lambda}}(\mathbf{z},
  {\mathbf{\tilde{z}}})
\end{equation}
with $z_j = z(x_j)$, $\tilde{z}_J = z(\tilde{x}_J)$,
\begin{equation}
\label{Psi0D}
  \Psi_0(\mathbf{x},{\mathbf{\tilde{x}}}) =
  \prod_{j=1}^N\psi_0(x_j)\prod_{J=1}^{\tilde{N}}
  \psi_{0,-1/\kappa}(\tilde{x}_J)\frac{\prod_{j<k}(z_k -
  z_j)^\kappa\prod_{J<K}(\tilde{z}_K - {\tilde
  z}_J)^{1/\kappa}}{\prod_{j,J}(\tilde{z}_J - z_j)},
\end{equation}
and where the $P_{\boldsymbol{\lambda}}$ are particular polynomials
contained in the algebra $\Lambda_{N,\tilde{N},\kappa}$. Moreover, the
corresponding eigenvalues are given by
\begin{equation}\label{EnD}
  E_{\boldsymbol{\lambda}} = E_0 -\sum_{j=1}^{\ell(\boldsymbol{\lambda})}
  \left(\alpha_2\lambda_j(\lambda_j - 1) + \left(2\alpha_2(\kappa(N -
  j) - \tilde{N}) + \beta_1\right)\lambda_j\right)
\end{equation}
with
\begin{multline}\label{E0D}
E_0 = -\frac{\kappa^2\alpha_2 }3\left( \left(N - \tilde{N}/\kappa
  \right)^3 - N + \tilde{N}/\kappa^3 \right) \\ - \frac{\kappa(\beta_1
  - (1+\kappa)\alpha_2)}2\left( \left(N - \tilde{N}/\kappa \right)^2 -
  N -\tilde{N}/\kappa^2 \right).
\end{multline}
To the best of our knowledge, this fact was previously known only in
special cases; see \cite{SV,SV2}.

\subsection{Construction of eigenfunctions}\label{sec41}
At this point we fix the polynomials $\alpha$ and $\beta$ in
\eqref{ab} and the coupling constant $\kappa>0$ and consider the
resulting deformed Calogero-Sutherland operator $H_{N,\tilde{N}}$
\eqref{D}. As we will see, the construction of explicit series
representations for the polynomials $P_{\boldsymbol{\lambda}}$ is very
similar to the one in the previous section, but there are a few
important differences in definitions and notation which we now
specify.

We start by noting that $\Psi_{\boldsymbol{\lambda}}$ is an
eigenfunction of $H_{N,\tilde{N}}$ if and only if
$P_{\boldsymbol{\lambda}}$ is an eigenfunction of the differential
operator
\begin{multline}
\label{fact} 
\tilde{H}_{N,\tilde{N}} := \Psi_0^{-1}(H_{N,\tilde{N}} - E_0)\Psi_0 =
-\sum_{j=1}^N \left(\alpha(z_j)\partial^2_{z_j} + \beta
(z_j)\partial_{z_j}\right) \\ + \kappa \sum_{J=1}^{\tilde{N}}
\left(\alpha(\tilde{z}_J)\partial^2_{\tilde{z}_J} +
\beta_{-1/\kappa}(\tilde{z}_J)\partial_{\tilde{z}_J}\right) - 2\kappa
\sum_{j\neq k}\frac{\alpha(z_k)}{z_j - z_k}\partial_{z_j} \\ + 2
\sum_{J\neq K}\frac{\alpha(\tilde{z}_K)}{\tilde{z}_J - \tilde
  z_K}\partial_{\tilde{z}_J} - 2\kappa \sum_{J,k}
\frac{\alpha(z_k)}{\tilde{z}_J - z_k}\partial_{\tilde{z}_J} - 2
\sum_{j,K}\frac{\alpha(\tilde{z}_K)}{z_j - \tilde{z}_K}\partial_{z_j}
\end{multline}
with corresponding eigenvalue $E_{\boldsymbol{\lambda}}-E_0$; this
equivalence follows from a straightforward computations using that
$\Psi_0$ \eqref{Psi0D} is an eigenstate of $H_{N,\tilde{N}}$ \eqref{D}
with eigenvalue $E_0$ (Corollary~\ref{cor4} for $M=\tilde{M}=0$) and
the identities \eqref{th}, \eqref{W}, and \eqref{R2}.  We shall refer
to the polynomials $P_{\boldsymbol{\lambda}}$ as reduced
eigenfunctions of the deformed Calogero-Sutherland operators
\eqref{D}. Furthermore, we let
\begin{multline}\label{tFNMNM}
\tilde{F}_{N,\tilde{N},M,\tilde{M}}(\mathbf{z}, \mathbf{\tilde{z}},
\mathbf{w}, \mathbf{\tilde{w}})\\ = \frac{
\prod_{j<k}(1-w_j/w_k)^{\kappa} \prod_{J<K}
(1-\tilde{w}_J/\tilde{w}_K)^{1/\kappa}} {\prod_{j,J}
(1-w_j/\tilde{w}_J)}\\ \times \frac{\prod_{j,K}(1
-z_j/\tilde{w}_K)\prod_{J,k}(1 -\tilde{z}_J/w_k)}
{\prod_{j,k}(1-z_j/w_k)^\kappa \prod_{J,K}(1-\tilde{z}_J /
\tilde{w}_K)^{1/\kappa}}
\end{multline}
where 
\begin{equation*}
(\mathbf{z},\mathbf{\tilde{z}})=(z_{1},\ldots,z_{N},\tilde{z}_{1},
\ldots, \tilde{z}_{\tilde{N}})\; \mbox{ and }\;
(\mathbf{w},\mathbf{\tilde{w}})=(w_1,\ldots,w_N,\tilde{w}_1, \ldots,
\tilde{w}_{\tilde{M}})
\end{equation*}
and define the polynomials $f^{(M,\tilde{M})}_{\mathbf{n}}(
\mathbf{z}, \mathbf{\tilde{z}})$, $\mathbf{n}\in
\mathbb{Z}^{M+\tilde{M}}$, through the following expansion:
\begin{multline}\label{fMM}
\tilde{F}_{N,\tilde{N},M,\tilde{M}} (\mathbf{z}, \mathbf{\tilde{z}},
  \mathbf{w}, \mathbf{\tilde{w}})\\ = \sum_{\mathbf{n}
  \in\mathbb{Z}^{M+\tilde{M}}} f^{(M,\tilde{M})}_{ \mathbf{n}}
  (\mathbf{z}, \mathbf{\tilde{z}}) w_1^{-n_1} \cdots
  w_M^{-n_M}\tilde{w}_1^{-n_{M+1}} \cdots
  \tilde{w}_{\tilde{M}}^{-n_{M+\tilde{M}}},
\end{multline}
valid in the region $|w_1|>\cdots >|w_M|>|\tilde{w}_1|>\cdots >
|\tilde{w}_{\tilde{M}}|>\max_{j,J}(|z_j|,|\tilde{z}_J|)$. The next
step is to compute the action of the differential operator
$\tilde{H}_{N,\tilde{N}}$ on these polynomials
$f^{(M,\tilde{M})}_{\mathbf{n}}$. In order to simplify such a
computation it is useful to introduce the following `parity' function
$q$ on the index set $\lbrace 1,\ldots,M+\tilde{M}\rbrace$:
\begin{equation}
\label{qj}
  q(j) :=
  \begin{cases}
    0, & \text{if}~j = 1,\ldots,M,\\
    1, & \text{if}~j = M + 1,\ldots,\tilde{M}.
  \end{cases}
\end{equation}
We also find it convenient to define the following shifted integer
vectors $\mathbf{n}^+$ associated with each quantum number
$\mathbf{n}$:
\begin{equation}\label{npjD}
  n_j^+ =
  \begin{cases}
    n_j + \kappa(N + 1 - j) - \tilde{N}, & \text{if}~j =
    1,\ldots,M,\\ n_j + (\tilde{N} + M + 1 - j)/\kappa - N + M, &
    \text{if}~j = M + 1,\ldots,\tilde{M}.
  \end{cases}
\end{equation}
Proceeding in analogy with the proof of Lemma \ref{actionLemma} in the
previous section it is now straightforward to obtain the following:

\begin{lemma}\label{actionLemmaD}
For each $\mathbf{n}\in\mathbb{Z}^{M+\tilde{M}}$,
\begin{multline}\label{actionD}
  \tilde{H}_{N,\tilde{N}}f^{(M,\tilde{M})}_{\mathbf{n}} = \left(
  E_{\mathbf{n}}^{(M,\tilde{M})} -
  E_0\right)f^{(M,\tilde{M})}_{\mathbf{n}}\\ -
  \sum_{j=1}^{M+\tilde{M}} \left((-\kappa)^{q(j)}\alpha_1 n_j^+(n_j^+
  - 1) + \left(\beta_0 - \left(1 -
  (-\kappa)^{1-q(j)}\right)\alpha_1\right)(n_j^+ -
  1)\right)f^{(M,\tilde{M})}_{\mathbf{n}- \mathbf{e}_j}\\ -
  \alpha_0\sum_{j=1}^{M+\tilde{M}}(-\kappa)^{q(j)}(n_j^+ - 1)(n_j^+ -
  2)f^{(M,\tilde{M})}_{\mathbf{n}-2 \mathbf{e}_j}\\ + \sum_{j<k}(1 -
  \kappa)(-\kappa)^{1-q(j)-q(k)}
  \sum_{p=0}^2\sum_{\nu=1}^\infty\alpha_p(2\nu -
  p)f^{(M,\tilde{M})}_{\mathbf{n}-\mathbf{E}^{p,\nu}_{j,k}}
\end{multline}
with
\begin{multline}\label{EnD1}
E^{(M,\tilde{M})}_{\mathbf{n}} -E_0 = - \sum_{j=1}^M
\big(\alpha_2n_j(n_j-1) + (2\alpha_2( \kappa (N-j) -\tilde{N}) +
\beta_1)n_j\big) \\ + \sum_{j=M+1}^{M+\tilde{M}} \big(\kappa \alpha_2
n_j (n_j +1) + (2\alpha_2(\tilde{N}-M +1- j -\kappa (N-M))-\beta_1)
n_j\big).
\end{multline}
\end{lemma} 

Similarly to the previous section, Lemma \ref{actionLemmaD} implies
that that there exist an eigenfunction of the form
\begin{equation}\label{eigFuncAnsatzD}
  P^{(M,\tilde{M})}_{\mathbf{n}} = f^{(M,\tilde{M})}_{\mathbf{n}} + \sum_{\mathbf{m}}u_{\mathbf{n}}(\mathbf{m})f^{(M,\tilde{M})}_{\mathbf{m}},
\end{equation}
where the sum is over integer vectors
$\mathbf{m}\in\mathbb{Z}^{M+\tilde{M}}$ such that
$\mathbf{m}\prec\mathbf{n}$, if and only if the coefficients
$u_{\mathbf{n}}(\mathbf{m})$ satisfy the recursion relation
\begin{multline*}
	\left(E^{(M,\tilde{M})}_{\mathbf{n}} - E^{(M,\tilde{M})}_{\mathbf{m}}\right)u_{\mathbf{n}}(\mathbf{m})\\ = -
  \sum_{j=1}^{M+\tilde{M}} m^+_j\left((-\kappa)^{q(j)}\alpha_1 (n_j^+
  + 1) + \left(\beta_0 - \left(1 -
  (-\kappa)^{1-q(j)}\right)\alpha_1\right)\right)u_{\mathbf{n}}(\mathbf{m}+e_j)\\ -
  \alpha_0\sum_{j=1}^{M+\tilde{M}}(-\kappa)^{q(j)}m_j^+(m_j^+ +
  1)u_{\mathbf{n}}(\mathbf{m}+2e_j)\\ + (1 -
  \kappa)\sum_{j<k}(-\kappa)^{1-q(j)-q(k)}
  \sum_{p=0}^2\sum_{\nu=1}^\infty\alpha_p(2\nu -
  p)u_{\mathbf{n}}(\mathbf{n}+\mathbf{E}^{p,\nu}_{j,k}).
\end{multline*}
Provided that $E^{(M,\tilde{M})}_{\mathbf{n}} -
E^{(M,\tilde{M})}_{\mathbf{m}}\neq 0$ for all integer vectors
$\mathbf{m}$ which appear in \eqref{eigFuncAnsatzD} this uniquely
determines all coefficients $u_{\mathbf{n}}(\mathbf{m})$. As
previously mentioned, an important feature of our construction is that
the non-negative integers $M$ and $\tilde{M}$ can be chosen
freely, and for each choice we obtain a family of reduced
eigenfunctions $P^{(M,\tilde{M})}_{\mathbf{n}}$ of $H_{N,\tilde{N}}$
which are labelled by integer vectors
$\mathbf{n}\in\mathbb{Z}^{M+\tilde{M}}$. This yields eigenfunctions
\eqref{PsinD} of $H_{N,\tilde{N}}$ with corresponding eigenvalues
$E^{(M,\tilde{M})}_{\mathbf{n}}$ given by \eqref{EnD1} and
\eqref{E0D}. A priori, it is not clear whether these eigenvalues
coincide with the ones given in \eqref{EnD}. In particular, it seems
that the eigenvalues we obtain depend on the specific values of $M$
and $\tilde{M}$, but, of course, the spectrum of $H_{N,\tilde{N}}$
does not. This apparent paradox can be resolved by exhibiting an
explicit mapping from the eigenvalues $E^{(M,\tilde{M})}_{\mathbf{n}}$
\eqref{EnD1} to $E_{\boldsymbol{\lambda}}$ \eqref{EnD} under certain
restrictions on the integer vectors $\mathbf{n}\in
\mathbb{Z}^{M+\tilde{M}}$.  To state this result we need the notion of
the conjugate ${\boldsymbol{\mu}}^\prime$ of a partition
${\boldsymbol{\mu}}$, obtained by interchanging rows and columns in
the Young diagram corresponding to ${\boldsymbol{\mu}}$, e.g.,
\begin{equation*}
  {\boldsymbol{\mu}}=(5,3,2):\quad \mbox{\tiny \Yvcentermath1 $\yng(5,3,2)$}
  \to \mbox{\tiny\Yvcentermath1 $\yng(3,3,2,1,1)$}:\quad {\boldsymbol{\mu}}' =
  (3,3,2,1,1);
\end{equation*}
see, e.g., Section~I.1 in \cite{MacD}.

\begin{lemma}\label{EnLemma}
Let $\mathbf{n} = (\mathbf{m},{\boldsymbol{\mu}})$ with $\mathbf{m} \in
\mathbb{Z}^M$ and ${\boldsymbol{\mu}}$ a partition of length
$\ell({\boldsymbol{\mu}})\leq \tilde{M}$. Then
\begin{equation}
\label{EE}
  E^{(M,\tilde{M})}_{\mathbf{n}} = E_{\boldsymbol{\lambda}}
\end{equation}
with $\boldsymbol{\lambda} = (\mathbf{m},{\boldsymbol{\mu}}^\prime)$.
\end{lemma}

\begin{proof}
We recall the following two well-known identities obeyed by all
partitions ${\boldsymbol{\mu}}$:
\begin{equation*}
  \sum_j\mu_j^2 = \sum_j (2j-1)\mu_j',\quad \sum_j j \mu_j =
  \sum_j\frac{1}{2}\mu_j'(\mu_j'+1),
\end{equation*}
where the sums are over all non-zero parts $\mu_j$ and $\mu^\prime_j$
of ${\boldsymbol{\mu}}$ and ${\boldsymbol{\mu}}^\prime$,
respectively. This, together with the identity $\sum_j\mu_j =
\sum_j\mu^\prime_j$ inserted into the second sum in \eqref{EnD1},
yields
\begin{multline*}
  E^{(M,\tilde{M})}_{\mathbf{n}} = E_0
  -\sum_{j=1}^M\big(\alpha_2m_j(m_j-1) +(2\alpha_2( \kappa (N-j)
  -\tilde{N}) +\beta_1)m_j\big) \\ -
  \sum_{j=1}^{\ell(\boldsymbol{\mu}')}\big(\alpha_2\mu_j'(\mu_j'-1) + (2
  \alpha_2(\kappa (N-M- j)-\tilde{N})+ \beta_1) \mu'_j\big).
\end{multline*}
The statement now follows by substituting $\lambda_j$ for $m_j$ and
$\lambda_{M+j}$ for $\mu'_j$.
\end{proof}

\subsection{On the complexity of the series representations}
\label{sec42}
The construction described in the previous section gives reduced
eigenfunctions $P_{\mathbf n}$ for all integer vectors $\mathbf{n}$
such that a certain condition of non-degenracy on the corresponding
eigenvalues is satisfied, but unless $\mathbf{n}$ satisfies the
condition in \eqref{EE} for some partition $\boldsymbol{\lambda}$ in
the fat $(N,\tilde{N})$-hook this function ought to vanish
identically. Lemma~\ref{EnLemma} gives, for each partition
$\boldsymbol{\lambda}$ in the fat $(N,\tilde{N})$-hook, such an
integer vector $\mathbf{n}$ provided one chooses $(M,\tilde{M}) =
(N,\tilde{N})$, and the latter choice thus is natural if one is
interested in a general formula which works for (essentially) all
eigenfunctions. However, we stress that we have constructed a family
of reduced eigenfunctions for each choice of non-negative integers $M$
and $\tilde{M}$, and that there exist a simple relation between any
two such families: For any two sets of non-negative integers
$(M_1,\tilde{M}_1)$ and $(M_2,\tilde{M}_2)$ one can show that two
reduced eigenfunctions $P^{(M_1,\tilde{M}_1)}_{\mathbf{n}_{1}}$ and
$P^{(M_2,\tilde{M}_2)}_{\mathbf{n}_{2}}$, labelled by integer vectors
$\mathbf{n}_{1}$ and $\mathbf{n}_{2}$ which correspond to one and the
same partition $\boldsymbol{\lambda}$ under the mapping defined in
Lemma~\ref{EnLemma}, are equal up to normalisation; see the first
arXiv version (v1) of \cite{Hal2}. We can thus, in a simple manner,
obtain different series representations for one and the same reduced
eigenfunction by varying the values of $M$ and $\tilde{M}$. Since the
complexity of this series representation is highly dependent on the
values of $M$ and $\tilde{M}$, one can choose the latter such that the
complexity is minimised, in many cases substantially below that of the
canonical choice $(M,\tilde{M}) = (N,\tilde{N})$. This is already
evident in the case of the Schr\"odinger operators \eqref{HN}, i.e.,
for $\tilde{N} = 0$. For example, suppose that we are interested in
the reduced eigenfunction of the standard Sutherland model for $N = 8$
and corresponding to the partition with diagram
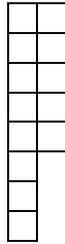
\begin{equation*}
  \yng(2,2,2,2,2,1,1,1) . 
\end{equation*}
Rather than setting $M = 8$ and $\tilde{M} = 0$ we can use
the fact that the conjugate of this partition is given by the diagram
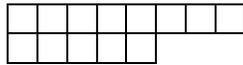
\begin{equation*}
  \yng(8,5), 
\end{equation*}
and thus the sought after reduced eigenfunction can equally well be
obtained by setting $M = 0$ and $\tilde{M} = 2$, in the process
decreasing the complexity of its series representation from the
$8$-particle case to that of only $2$-particles. As a further
illustrative example we consider the reduced eigenfunction
corresponding to the partition with diagram
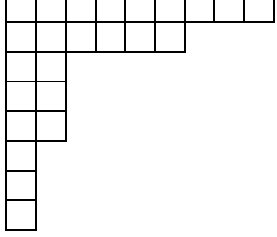
\begin{equation*}
  \yng(9,6,2,2,2,1,1,1)
\end{equation*}
for which we can set $M = \tilde{M} = 2$ to minimise the complexity of
its series representation. Similar examples can be easily constructed
also in the `deformed' case. In generally, it can be readily verified
that, given a specific partition $\boldsymbol{\lambda}$, the minimal
value for $M + \tilde{M}$ is attained by setting $M = j$ and
$\tilde{M} = \lambda_{j+1}$ where $j$ denotes the row in the diagram
of $\boldsymbol{\lambda}$ for which $j + \lambda_{j+1}$ is
minimal. This means, in particular, that we always can choose $M$ and
$\tilde{M}$ such that their sum does not exceed the length
$\ell(\boldsymbol{\lambda})$ of the partition $\boldsymbol{\lambda}$
in question, which may be less than $N + \tilde{N}$. This observation
is also reflected in the fact that $f_{\mathbf{n}}^{(M,\tilde{M})} =
f_{\mathbf{n}}^{(M,\tilde{M}+K)}$ for all non-negative values of $K$
as long as $M + \tilde{M}\geq\ell(\mathbf{n})$; the latter is readily
inferred from the definition of the polynomials
$f^{(M,\tilde{M})}_{\mathbf{n}}$.

\subsection{Elementary properties of the reduced eigenfunctions}
We conclude this section by verifying that the functions
$f^{(M,\tilde{M})}_{\mathbf{n}}$, and consequently also the reduced
eigenfunctions $P^{(M,\tilde{M})}_{\mathbf{n}}$, are contained in the
algebra $\Lambda_{N,\tilde{N},\kappa}$ for all values of the
non-negative integers $M$ and $\tilde{M}$.

\begin{lemma}
Let $\mathbf{n}\in\mathbb{Z}^{M+\tilde{M}}$. Then the function
$f^{(M,\tilde{M})}_{\mathbf{n}}$ is non-zero only if
$\mathbf{n}\succeq 0$, and in that case it is a homogeneous polynomial
of degree $|\mathbf{n}|$ in $\Lambda_{N,\tilde{N},\kappa}$.
\end{lemma}

\begin{proof}
The proof is by straightforward computations: expanding each factor on
the r.h.s.\ in \eqref{tFNMNM} in the region $|w_1| > \cdots >
|w_{M+\tilde{M}}|> \max_{j,J}(|z_j|,|\tilde{z}_J|)$ (recall that
$\tilde{w}_J = w_{M+J}$) in a binomial series one obtains a
well-defined series representation of
$\tilde{F}_{N,\tilde{N},M,\tilde{M}}$ as a linear superposition of
monomials $\mathbf{w}^{-\mathbf{n}}$. By inspection one can check that
in this series only terms $\mathbf{w}^{-\mathbf{n}}$ with
$\mathbf{n}\succeq 0$ appear, that each such term appear only a finite
number of times, and that the coefficients of
$\mathbf{w}^{-\mathbf{n}}$ is a homogeneous polynomials of degree
$|\mathbf{n}|$ in the variables $z_j$ and $\tilde{z}_J$. It thus
follows from \eqref{fMM} that $f^{(M,\tilde{M})}_{\mathbf{n}}$ is
non-zero only if $\mathbf{n}\succeq 0$, and in that case it is a
homogeneous polynomial of degree $|\mathbf{n}|$. To prove that the
$f^{(M,\tilde{M})}_{\mathbf{n}}$ are contained in
$\Lambda_{N,\tilde{N},\kappa}$ we observe that
$\tilde{F}_{N,\tilde{N},M,\tilde{M}}$ is separately symmetric in the
variables $z_j$ and $\tilde{z}_J$, and that
\begin{equation*}
  (\partial_{z_j}+\kappa\partial_{\tilde{z}_J})
  \tilde{F}_{N,\tilde{N},M,\tilde{M}}\big\rvert_{z_j=\tilde{z}_J}=0
\end{equation*}
for all $j = 1,\ldots,N$ and $J = 1,\ldots,\tilde{N}$.  The latter
follows from \eqref{tFNMNM} by straightforward computations,
\begin{equation*}
  \partial_{z_j} \log \tilde{F}_{N,\tilde{N},M,\tilde{M}} =
  \sum_{K=1}^{\tilde{M}} \frac1{z_j-\tilde{w}_K} - \kappa \sum_{k=1}^M
  \frac1{z_j- w_k}
\end{equation*}
and
\begin{equation*}
  \partial_{\tilde{z}_J} \log \tilde{F}_{N,\tilde{N},M,\tilde{M}} =
  \sum_{k=1}^M \frac1{\tilde{z}_J-w_k} -
  \frac1{\kappa}\sum_{K=1}^{\tilde{M}} \frac1{\tilde{z}_J-
  \tilde{w}_K}.
\end{equation*}
\end{proof}

\section{Concluding remarks}\label{sec5}
In this final section we briefly discuss the relation of our
construction to the so called generalised classical orthogonal
polynomials of Lassalle and Macdonald, present some further remarks of
the effect of spectral degeneracies on our constructions of reduced
eigenfunctions of the Schr\"odinger operators \eqref{HN} and the
deformed Calogero-Sutherland operators \eqref{D}, and comment on the
question of completeness in the case of the deformed
Calogero-Sutherland operators.

\subsection{The relation to generalised classical orthogonal polynomials}
As is well known, the spectra of the Schr\"odinger operators
\eqref{HN} are degenerate. In particular, in cases I and IV the
eigenvalue of a given reduced eigenfunction $P_{\mathbf{n}}$ depends
only on the weight $|\mathbf{n}|$ of the integer vector
$\mathbf{n}$. An interesting question is thus what the relation is
between the basises for the reduced eigenfunctions we construct and
previously known such basises. Below we will sketch a proof of the
fact that our basis coincide, up to normalisation, with the
generalised classical orthogonal polynomials of Lassalle
\cite{Las,Las3,Las2} and Macdonald \cite{MacD2}. We recall that these
latter polynomials are reduced eigenfunctions of the Schr\"odinger
operator in case I (Hermite), case IV (Laguerre) or case VI (Jacobi),
and of the form
\begin{equation*}
  \mathcal{P}_{\boldsymbol{\lambda}} = \sum_{\boldsymbol{\mu}\subseteq\boldsymbol{\lambda}}u_{\boldsymbol{\lambda}\boldsymbol{\mu}}\mathcal{J}_{\boldsymbol{\mu}}
\end{equation*}
for some coefficients $u_{\boldsymbol{\lambda}\boldsymbol{\mu}}$, and
where $\boldsymbol{\mu}\subseteq\boldsymbol{\lambda}$ means that the
Young diagram of $\boldsymbol{\mu}$ is contained in that of
$\boldsymbol{\lambda}$. We shall require the following partial order
on the set of partitions:
\begin{equation*}
  \boldsymbol{\mu}\leq\boldsymbol{\lambda} \Leftrightarrow \mu_1+\cdots+\mu_j\leq\lambda_1+\cdots+\lambda_j,\quad \forall j.
\end{equation*}
When restricted to the set of partitions of a fixed weight this
coincides with the so called dominance order. It is a simple exercise
to verify that if $|\boldsymbol{\mu}| = |\boldsymbol{\lambda}|$ and
$\boldsymbol{\mu}\leq\boldsymbol{\lambda}$ then
$\boldsymbol{\mu}\succeq\boldsymbol{\lambda}$ and vice
versa. Furthermore, comparing the expansions \eqref{PiExpInJacks} and
\eqref{PiExpInGs}, and using the triangular structure of the Jack
polynomials when expanded in monomial symmetric polynomials, one finds
that
\begin{equation*}
  g_{\boldsymbol{\lambda}} = \sum_{|\boldsymbol{\mu}| = |\boldsymbol{\lambda}|,\boldsymbol{\mu}\geq\boldsymbol{\lambda}} a_{\boldsymbol{\lambda}\boldsymbol{\mu}}\mathcal{J}_{\boldsymbol{\mu}}
\end{equation*}
for some coefficients $a_{\boldsymbol{\lambda}\boldsymbol{\mu}}$ such
that $a_{\boldsymbol{\lambda}\boldsymbol{\lambda}} =
b_{\boldsymbol{\lambda}}$; see Equation 10.16 in Section VI.10 of
\cite{MacD} for the definition of $b_{\boldsymbol{\lambda}}$ (in doing
so observe that the parameter $\alpha = 1/\kappa$). It is now readily
inferred from Lemma \ref{ftriangLemma} that any reduced eigenfunction
$P_{\boldsymbol{\lambda}}$, as constructed in Section \ref{sec31}, is
of the form
\begin{equation*}
  P_{\boldsymbol{\lambda}} = \sum_{\boldsymbol{\mu}\preceq\boldsymbol{\lambda}}v_{\boldsymbol{\lambda}\boldsymbol{\mu}}\mathcal{J}_{\boldsymbol{\mu}}
\end{equation*}
for some coefficients $v_{\boldsymbol{\lambda}\boldsymbol{\mu}}$ such
that $v_{\boldsymbol{\lambda}\boldsymbol{\lambda}} =
b_{\boldsymbol{\lambda}}$. Since
$\boldsymbol{\mu}\preceq\boldsymbol{\lambda}$ if
$\boldsymbol{\mu}\subseteq\boldsymbol{\lambda}$ and, up to
normalisation, there can be only one eigenfunction of this form, it
follows that the relation between the reduced eigenfunctions we
construct and those of Lassalle and Macdonald is given by
\begin{equation*}
  P_{\boldsymbol{\lambda}} = b_{\boldsymbol{\lambda}}\mathcal{P}_{\boldsymbol{\lambda}}.
\end{equation*}

\subsection{Spectral degeneracies}
We stress once more the issue of degeneracies of the eigenvalues of
the Schr\"odinger operators \eqref{HN} and the deformed
Calogero-Sutherland operators \eqref{D}: since our construction of
reduced eigenfunction of these operators only is valid under a certain
non-degeneracy condition on the corresponding eigenvalues, this issue
plays a decisive role in our construction. From our point of view, it
is essentially only this issue which truly distinguishes the different
special cases of our results listed in Table~\ref{table1}. It thus
would be interesting to investigate this issue further.

\subsection{Completeness for deformed Calogero-Sutherland operators}
We mention that one can prove an analogue of Proposition \ref{prop32}
for our construction of reduced eigenfunctions of the deformed
Calogero-Sutherland operators \eqref{D}. In fact, proceeding similarly
to Section \ref{sec33} one can prove that the reduced eigenfunctions
constructed in Section \ref{sec41} have a triangular expansion in
so-called super Jack polynomials. We recall that the super Jack
polynomials are known to constitute a linear basis for the algebra
$\Lambda_{N,\tilde{N},\kappa}$. Using these two facts above one can
show that if $M\geq N$ and $\tilde{M}\geq \tilde{N}$ then the reduced
eigenfunctions constructed in Section \ref{sec41} are also a linear
basis for $\Lambda_{N,\tilde{N},\kappa}$. If either $M<N$ or
$\tilde{M}<\tilde{N}$ then they span only a subspace of
$\Lambda_{N,\tilde{N},\kappa}$ which, however, can be given a rather
simple characterisation. A detailed discussion of these facts can be
found in the first arXiv version (v1) of \cite{Hal2}.

\section*{Acknowledgements} 
We are grateful to F.\ Calogero, O.\ Chalykh, A.\ Fordy, P.\
Forrester, Y. Suris, and A.\ Veselov, for useful discussions.  M.H.\
would also like to thank P.\ Forrester for making available to him an
unpublished manuscript by Macdonald \cite{MacD2}.  We are also
grateful to two anonymous referees for constructive criticism which
helped us to improve the paper. This work was significantly influenced
by discussions with V.\ Kuznetsov. His experience, critical remarks,
and interest were very valuable to us and are now sadly missed. This
work was supported by the Swedish Science Research Council (VR), the
G\"oran Gustafsson Foundation, and the European Union through the FP6
Marie Curie RTN {\em ENIGMA} (Contract number MRTN-CT-2004-5652).

\end{document}